\documentclass[aps,prd,showpacs,notitlepage,nofootinbib,preprintnumbers,amsmath,amssymb]{revtex4-1}

\newcommand{\tr}{\mbox{tr}}
\newcommand{\thalf}{\tfrac{1}{2}}
\newcommand{\llangle}{\Big\langle \!\! \Big\langle}
\newcommand{\rrangle}{\Big\rangle \!\! \Big\rangle}

\usepackage{graphics,graphicx}
\usepackage{dcolumn}
\usepackage{bm}
\usepackage{epsfig}
\usepackage[usenames]{color}
\usepackage{hyperref} 
\usepackage{mathbbol}
\usepackage{epstopdf}
\usepackage{simplewick}
\usepackage[utf8]{inputenc} 
\usepackage{slashed}

\def\eq#1{{Eq.~(\ref{#1})}}
\def\fig#1{{Fig.~\ref{#1}}}
\newcommand{\ben}{\begin{eqnarray*}}
\newcommand{\een}{\end{eqnarray*}}

\newcommand{\pd}{\partial}

\newcommand{\as}{\alpha_s}

\newcommand{\dhd}{{\textstyle d}
\lower.03ex\hbox{\kern-0.38em$^{\scriptstyle-}$}\kern-0.05em{}}
\newcommand{\dbar}{{\textstyle \delta}
\lower.03ex\hbox{\kern-0.38em$^{\scriptstyle-}$}\kern-0.05em{}}
\newcommand{\half}{{1\over 2}}

\newcommand{\ul}[1]{\underline{#1}}

\newcommand{\ord}[1]{\mathcal{O}\left( #1 \right)}

\begin{document}

\title{Helicity Evolution at Small $x$: Flavor Singlet and Non-Singlet Observables}
\author{Yuri V. Kovchegov} 
         \email[Email: ]{kovchegov.1@osu.edu}
         \affiliation{Department of Physics, The Ohio State
           University, Columbus, OH 43210, USA}
         \author{Daniel Pitonyak} \email[Email:
         ]{dpitonyak@quark.phy.bnl.gov} \affiliation{Division of Science, Penn
           State University-Berks, Reading, PA 19610, USA} \affiliation{RIKEN BNL
           Research Center, Brookhaven National Laboratory, Upton, New
           York 11973, USA}
\author{Matthew D. Sievert}
	\email[Email: ]{sievertmd@lanl.gov}
	\affiliation{Theoretical Division, Los Alamos National 
	        Laboratory, Los Alamos, NM 87545, USA}
	\affiliation{Physics Department, Brookhaven National
          Laboratory, Upton, NY 11973, USA \\}
\date{\today}%

\begin{abstract}
  We extend our earlier results for the quark helicity evolution at
  small $x$ \cite{Kovchegov:2015pbl} to derive the small-$x$
  asymptotics of the flavor singlet and flavor non-singlet quark
  helicity TMDs and PDFs and of the $g_1$ structure function. In the
  flavor singlet case we re-derive the evolution equations obtained in
  our previous paper on the subject \cite{Kovchegov:2015pbl},
  performing additional cross-checks of our results. In the flavor
  non-singlet case we construct new small-$x$ evolution equations by
  employing the large-$N_c$ limit. All evolution equations resum
  double-logarithmic powers of $\as \, \ln^2 (1/x)$ in the
  polarization-dependent evolution along with the single-logarithmic
  powers of $\as \, \ln (1/x)$ in the unpolarized evolution which
  includes saturation effects. We solve the linearized flavor
  non-singlet equation analytically, obtaining an intercept which
  agrees with the one calculated earlier by Bartels, Ermolaev and Ryskin
  \cite{Bartels:1995iu} using the infra-red evolution equations. Our
  numerical solution of the linearized large-$N_c$ evolution equations
  for the flavor singlet case is presented in the accompanying Letter
  \cite{Letter} and is further discussed here.
\end{abstract}


\maketitle

\section{Introduction}

Measurements of hadronic structure functions in deep inelastic
scattering are kinematically limited to a minimum value of Bjorken-$x$
due to a finite center-of-mass energy $s \propto \tfrac{1}{x}$.
Therefore, all structure functions, and all parton distribution
functions (PDF's) $f(x,Q^2)$ extracted from them, must necessarily be
extrapolated toward smaller $x$ in order to generate predictions for
higher energy scattering experiments and to apply Quantum
Chromodynamics (QCD) sum rules which constrain moments $\int_0^1 dx \,
x^n f(x,Q^2)$ of the PDF's.  Structure functions in the $x \rightarrow
0$ limit are often singular, with the best-known examples being the
unpolarized structure functions $F_1$ and $F_2$.  At leading twist and
leading order in the coupling $\as$, $F_1$ is a weighted measure of
the total density of partons in a hadron, with its $x \rightarrow 0$
singularity reflecting, in part, the enhancement of soft gluon
radiation in QCD.  The dynamics of this soft gluon radiation are
encapsulated in the Balitsky--Fadin--Kuraev--Lipatov (BFKL)
\cite{Kuraev:1977fs,Balitsky:1978ic}, Balitsky--Kovchegov (BK)
\cite{Balitsky:1996ub,Balitsky:1998ya,Kovchegov:1999yj,Kovchegov:1999ua},
and Jalilian-Marian--Iancu--McLerran--Weigert--Leonidov--Kovner
(JIMWLK) evolution equations
\cite{Jalilian-Marian:1997dw,Jalilian-Marian:1997gr,Iancu:2001ad,Iancu:2000hn},
which describe the development of a cascade of small-$x$ gluons at
high energies by resumming (at leading order) the large logarithms
$\alpha_s \ln\tfrac{s}{\Lambda^2} \sim \alpha_s \ln\tfrac{1}{x} \sim
1$ ($\Lambda$ is an infrared cutoff).  This parton cascade, as
described by the linear BFKL equation, leads to a steep growth in the
gluon number density $\tfrac{d N}{dy} \propto x f(x, Q^2)$ of partons
per unit rapidity resulting in a violation of the black disk limit for
the corresponding scattering cross sections.  This growth is regulated
by the onset of the high-density regime of QCD, where nonlinear
multiple rescatterings of the parton cascade in the target (as
described by the BK and JIMWLK equations) {\textit{saturate}} the
number density of partons such that $dN/dy$ remains finite as $x\to
0$. This results in corresponding cross sections satisfying the black
disk limit (see
\cite{Gribov:1984tu,Iancu:2003xm,Weigert:2005us,Jalilian-Marian:2005jf,Gelis:2010nm,Albacete:2014fwa,KovchegovLevin}
for reviews).

The BFKL, BK, and JIMWLK equations, however, cannot describe the
small-$x$ limit of the {\it{polarized}} structure function $g_1$,
which at leading twist and leading order in $\as$ is a weighted
measure of the quark helicity PDF's $\Delta q^f (x, Q^2)$.  The
high-energy / small-$x$ asymptotics captured by these evolution
equations are insensitive to polarization because, as is well-known,
polarization dependence is suppressed at high energies.  The
sub-eikonal interactions which do not transfer spin in high-energy
scattering enter as power-suppressed corrections to unpolarized
cross-sections and to $F_1$ and $F_2$.  The sub-eikonal interactions
which do transfer longitudinal spin provide the {\textit{leading}}
high-energy / small-$x$ asymptotics of the longitudinal double-spin
asymmetry $A_{LL}$ and the polarized structure function $g_1$.  The
development of a cascade of {\textit{polarized}} partons at small $x$
is thus an interesting and important aspect of high-energy dynamics in
QCD which is outside the scope of the canonical small-$x$ treatment.

The small-$x$ asymptotics of the polarized $g_1$ structure function
were studied previously by Bartels, Ermolaev, and Ryskin (BER) in the
flavor singlet \cite{Bartels:1996wc} and non-singlet
\cite{Bartels:1995iu} cases.  Unlike with unpolarized small-$x$
evolution, in helicity evolution, $t$-channel quarks play an important
role already at the leading order.  In the massless limit, quarks
automatically transfer spin through the $t$-channel due to helicity
conservation for massless fermions.  The effective particle exchanged
by such dressed quarks is known as the Reggeon in the small-$x$
literature \cite{Kirschner:1983di, Kirschner:1985cb, Kirschner:1994vc,
  Kirschner:1994rq, Griffiths:1999dj}, and it has been studied
previously in the context of baryon stopping in heavy-ion collisions
\cite{Itakura:2003jp}. In addition to quarks, a power-suppressed
component of gluon exchange can also carry spin through the
$t$-channel.  While the unpolarized BFKL / BK / JIMWLK evolution
occurs through the exchange of dressed, longitudinally-polarized
gluons (the ``hard QCD Pomeron''), helicity evolution receives
contributions from exchanging a pair of dressed gluons with one
polarized longitudinally and the other transversely
\cite{Bartels:1995iu,Bartels:1996wc}.  The exchange of quarks and of
polarized gluons both enter at the same parametric order and can
therefore mix with each other, akin to the mixing which occurs in
polarized and unpolarized
Dokshitzer--Gribov--Lipatov--Altarelli--Parisi (DGLAP) evolution in
$Q^2$ \cite{Dokshitzer:1977sg,Gribov:1972ri,Altarelli:1977zs}.

Helicity evolution (like Reggeon evolution) is double-logarithmic,
resumming two logarithms of the energy for each power of the coupling:
$\alpha_s \ln^2 \tfrac{s}{\Lambda^2} \sim \alpha_s \ln^2\tfrac{1}{x}
\sim 1$.  In this sense, helicity evolution can be said to be
{\textit{stronger}} than the single-logarithmic BFKL / BK / JIMWLK
evolution, leading to the possibility that the helicity PDF's could
become (almost) competitive with the unpolarized ones at small $x$,
despite their suppression in the initial conditions.  Indeed, this is
what BER found \cite{Bartels:1996wc}: for the flavor singlet case with
$N_f = 0$ (pure glue) and $\alpha_s (Q^2) = 0.343$ at $Q^2 = 3~GeV^2$,
their results required helicity PDF's to grow at small $x$ as
$(\tfrac{1}{x})^{1.481}$.  In comparison, leading-order fixed-coupling
BFKL evolution with the same parameters yields unpolarized PDF's which
grow at small $x$ as $(\tfrac{1}{x})^{1.908}$.  The presence of a
non-integrable singularity in the helicity PDF's would imply that
their contribution to the proton spin $S_q = \thalf \sum_f \int_0^1 dx
\, \Delta q^f (x, Q^2)$ is not finite, requiring either higher-order
corrections or nonlinear saturation effects at small $x$ to regulate
the divergence.  The latter scenario would potentially provide a novel
path to discovering parton saturation using measurements of the
polarized structure functions instead of the unpolarized ones.

Motivated by this possibility, and by the need to assess the amount of
proton spin at small $x$, we derived in a previous work
\cite{Kovchegov:2015pbl} evolution equations for the quark helicity
PDF's at small $x$, including the nonlinear multiple rescattering
which drives parton saturation.  Our approach used the modern
saturation formalism, relating the helicity PDF's to a polarized
dipole amplitude which we calculated in light-front perturbation
theory (LFPT) \cite{Lepage:1980fj}.  The resulting evolution equations
involve quark and gluon Wilson line operators, along with an object we
refer to as the ``polarized Wilson line": an eikonal quark or gluon
propagator with the insertion of one or two sub-eikonal vertices
carrying polarization information.  These equations do not close in
general because they involve higher-order operators in the evolution
kernel; this should result in a helicity analogue of Balitsky
hierarchy \cite{Balitsky:1996ub,Balitsky:1998ya}. However, also by
analogy to the unpolarized case, our helicity evolution equations do
close in the large-$N_c$ and large-$N_c \, \& N_f$ limits, with $N_c$
the number of colors and $N_f$ the number of flavors. The equations
are quite complex and difficult to solve, even in the linearized
strictly double-logarithmic regime.  In an accompanying Letter
\cite{Letter} we present the numerical solution of our equations at
large $N_c$, obtaining the helicity intercept of $\alpha_{h} \approx
2.31 \: {\sqrt{\tfrac{\alpha_s N_c}{2 \pi}}}$. This leads to helicity
PDF's with {\it{integrable}} singularities at small $x$, scaling as
$\Delta q \sim \left( \tfrac{1}{x} \right)^{\alpha_h} \sim \left(
  \tfrac{1}{x} \right)^{0.936}$ for $N_f = 0$ and $Q^2 = 3~GeV^2$, and
hence a finite value of the quark spin contribution $S_q$.  Such a
scenario would not, in fact, require saturation effects to regulate
the small-$x$ limit after all.

Somewhat surprisingly, the value of our helicity intercept in the
flavor singlet channel is smaller than that obtained by BER by about
$35\%$ \cite{Bartels:1996wc}. To understand the source of our
significant discrepancy with BER, we have performed a variety of
consistency checks of our equations, which we present here in detail.
This analysis also sheds further light on the intricate structure of
helicity evolution at small $x$, which is substantially more complex
than the unpolarized evolution which is well-known in the literature.

While our previous paper \cite{Kovchegov:2015pbl} dealt with
flavor-singlet helicity observables, here we have also generalized the
treatment to include flavor non-singlet helicity PDF's and transverse
momentum-dependent PDF's (TMD's), along with the $g_1$ structure
function. Constructing a large-$N_c$ helicity evolution equation for
the flavor non-singlet case, we have reproduced the flavor non-singlet
intercept obtained previously by BER \cite{Bartels:1995iu}.

This paper is organized as follows.  In Sec.~\ref{FS} we rederive,
cross-check and present a solution for the helicity evolution
equations in the flavor singlet case derived previously in
\cite{Kovchegov:2015pbl}. We relate the polarized flavor-singlet
observables to a ``polarized dipole amplitude'' which contains the
dynamics of spin exchange at small $x$ in Sec.~\ref{FSdef}, and we
state our initial conditions for this amplitude. The large-$N_c$
flavor-singlet helicity evolution equations are presented in
Sec.~\ref{sec:FSevol}; they are solved numerically in the accompanying
Letter \cite{Letter}. We discuss the solution in Sec.~\ref{sec:FSsol}
and outline our disagreement with BER. In Sec.~\ref{sec:Cancellations}
we perform a number of explicit calculations which elucidate the role
of virtual corrections in our evolution equations and which verify the
real-virtual cancellations used in deriving them.  In
Sec.~\ref{sec:DGLAP} we use our evolution equations to compute the
glue/glue next-to-leading order (NLO) anomalous dimension in polarized
DGLAP evolution, again obtaining agreement with the literature
\cite{Mertig:1995ny} and with BER on this point. The flavor
non-singlet evolution is constructed in Sec.~\ref{sec:FNS}, following
the same pattern. The flavor non-singlet observables are defined in
Sec.~\ref{FNSdef} in terms of the flavor non-singlet ``polarized
dipole amplitude". The helicity evolution equations in the flavor
non-singlet case and in the large-$N_c$ limit are derived in
Sec.~\ref{FNSevol}, and are solved analytically in Sec.~\ref{FNSsol},
leading to an intercept in perfect agreement with
\cite{Bartels:1995iu}.  In Sec.~\ref{sec:Conclusions} we conclude by
summarizing the importance of our calculation for assessing the
small-$x$ contribution to the spin puzzle.

\section{Flavor Singlet Helicity Evolution}
\label{FS}

\subsection{Definitions and Initial Conditions}
\label{FSdef}

\begin{figure}[htb]
\centering
\includegraphics[width= 0.9 \textwidth]{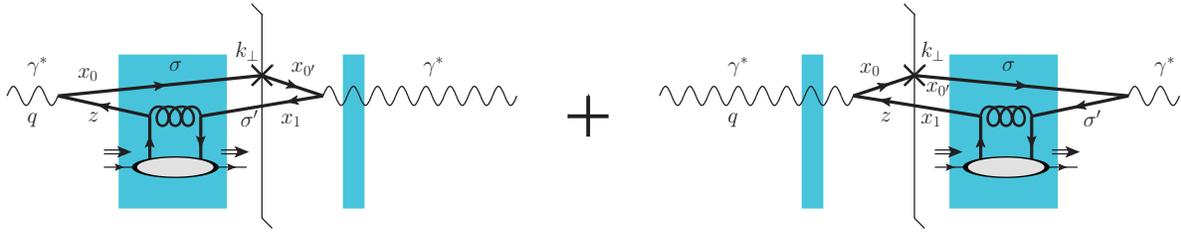}
\caption{Diagrams contributing to the quark helicity at
  small-$x$. The shaded region is the shock wave of the (polarized)
  target. The spin-dependent interaction is illustrated by $t$-channel
  quark exchanges, but in general should include gluon exchanges as
  well (see text).}
\label{f:helicity_TMD}
\end{figure}

As was derived in \cite{Kovchegov:2015zha,Kovchegov:2015pbl}, at small
$x$, the polarized structure function $g_1 (x , Q^2)$, the quark
helicity PDF $\Delta q (x, Q^2)$, and the quark helicity TMD $g_{1L}
(x, k_T^2)$ can all be expressed in the following way:
\begin{subequations} \label{e:observables}
  \begin{align} 
    g_1 (x, Q^2) &= \frac{N_c}{(2\pi)^2 \alpha_{EM}}
    \int\limits_{z_i}^1\frac{dz }{z^2 (1-z)} \, \int d x_{01}^2 \, d^2
    b \, \left[ \half \sum_{\lambda \sigma \sigma'} | \psi_{\lambda
        \sigma \sigma'}^T |^2_{(x_{01}^2 , z)} + \sum_{\sigma \sigma'}
      |\psi_{\sigma \sigma'}^L|^2_{(x_{01}^2 , z)} \right] \notag \\
    & \times \frac{1}{2 N_c} \left\{ \llangle \tr \left[ V_{\ul 0}
        V_{\ul 1}^{pol \, \dagger} \right] \rrangle (z) + \llangle \tr
      \left[ V_{\ul 0} V_{\ul 1}^{pol \, \dagger} \right] \rrangle^*
      (z) \right\} ,
  \\
  \Delta q(x, Q^2) &= \frac{N_c}{4 \pi^3} \int\limits_{z_i}^1
  \frac{dz}{z} \int\limits_{\tfrac{1}{z s}}^{\tfrac{1}{z Q^2}} \frac{d
    x_{01}^2}{x_{01}^2} \, \int d^2 b \: \frac{1}{2 N_c} \left\{
    \llangle \tr \left[ V_{\ul 0} V_{\ul 1}^{pol \, \dagger} \right]
    \rrangle (z) + \llangle \tr \left[ V_{\ul 0} V_{\ul 1}^{pol \,
        \dagger} \right] \rrangle^* (z) \right\} ,
  \\
  g_{1L} (x, k_T^2) &= \frac{4 N_c}{(2\pi)^6} \int\limits_{z_i}^1
  \frac{dz}{z} \int d^2 x_{01} \, d^2 x_{0'1} \: e^{-i \ul{k} \cdot
    (\ul{x}_{01} - \ul{x}_{0' 1})} \: \frac{\ul{x}_{01} \cdot
    \ul{x}_{0'1}}{x_{01}^2 x_{0'1}^2} \notag \\ & \times \, \int d^2
  b \: \frac{1}{2 N_c} \left\{ \llangle \tr \left[ V_{\ul 0} V_{\ul
        1}^{pol \, \dagger} \right] \rrangle (z) + \llangle \tr \left[
      V_{\ul 0} V_{\ul 1}^{pol \, \dagger} \right] \rrangle^* (z)
  \right\} . \label{g1L}
  \end{align}
\end{subequations}
These results come from the computation of the diagrams shown in
Fig.~\ref{f:helicity_TMD} in LFPT in the conventions of
\cite{Kovchegov:2012mbw}, where we take the virtual photon with
virtuality $Q^2$ to have a large momentum along the light-front
``$+$'' axis and work in the $A^+ = 0$ gauge. The diagrams in
Fig.~\ref{f:helicity_TMD} represent contributions to the
polarization-dependent part of the quark production cross section in
semi-inclusive deep inelastic scattering (SIDIS) on a polarized target
proton or nucleus; the quark helicity TMD and PDF and the $g_1$
structure function can be extracted from this quantity
\cite{Kovchegov:2015pbl}. The notation is defined as in
Fig.~\ref{f:helicity_TMD}: $\sigma , \sigma' , \lambda$ are the
polarizations of the quark, antiquark, and (transverse) photon,
respectively (we take the target to have positive helicity, without
loss of generality); $\ul{x}_1$ is the transverse coordinate of the
antiquark which scatters in a polarization-dependent way; $\ul{x}_0$
and $\ul{x}_{0'}$ are the transverse coordinates of the produced quark
in the amplitude and complex-conjugate amplitude, respectively; and
$z$ is the fraction of the photon's ``$+$'' momentum which is carried
by the antiquark.  Transverse vectors are denoted $\ul{v} \equiv
(v_\bot^1 , v_\bot^2)$, with $v_\perp = v_T \equiv |\underline{v}|$,
and the separation vector between coordinates is $\ul{x}_{i j} \equiv
\ul{x}_i - \ul{x}_j$. The dipole impact parameter is defined by
$\ul{b} = (\ul{x}_1 + \ul{x}_0)/2$. The $z$ integral has a lower
cutoff $z_i = \Lambda^2/s$ with $\Lambda$ the infrared (IR) cutoff and
$s$ the center-of-mass energy squared for the SIDIS process pictured
in \fig{f:helicity_TMD}. $N_c$ is the number of colors and
$\alpha_{EM}$ is the fine structure constant. The light-cone wave
functions for the $\gamma^* \to q \bar q$ splitting are denoted
$\psi_{\lambda \sigma \sigma'}^T$ and $\psi_{\sigma \sigma'}^L$ for
the transverse and longitudinal polarizations of the virtual photon
respectively. These functions are well-known in the literature
\cite{Bjorken:1970ah,Nikolaev:1990ja}, and are explicitly given
e.g. in \cite{Kovchegov:2015pbl}.

In obtaining the simplified expressions \eqref{e:observables} we have
taken the produced quarks to be massless and utilized parity symmetry
of the virtual photon wave functions. In arriving at \eq{g1L} we have
used the $\ul{k} \to - \ul{k}$ symmetry of the helicity TMD due to the
absence of any preferred transverse direction in the problem
\cite{Kovchegov:2015pbl}. The parton distribution functions in
Eqs.~\eqref{e:observables} are given to leading-twist accuracy, and
the structure function $g_1$ is given in the double-logarithmic
approximation. Note that so far we assume that only one (massless)
quark flavor enters the loop in the diagrams of \fig{f:helicity_TMD}.

The fundamental-representation Wilson line $V_{\ul 0}$ in
Eqs.~\eqref{e:observables} is the usual path-ordered exponential which
describes the gauge rotation of an eikonal quark passing through a
background gluon field of the target proton or nucleus:
\begin{align} 
  \label{e:Pexp}
  V_{\ul x} \equiv \mathcal{P} \exp \left[ i g
    \int\limits_{-\infty}^\infty d x^+ \, A^- (x^+ , 0^-, \ul{x})
  \right] .
\end{align}
Note the abbreviated notation $V_{\ul 0} \equiv V_{{\ul x}_0}$.

The ``polarized Wilson lines'' $V_{\ul x}^{pol}$ are more difficult to
define operatorially. Consider an eikonal quark propagator with the
insertion of either one (for gluon exchange) or two (for quark
exchange) sub-eikonal polarization-dependent vertices. The resulting
propagator of an eikonal quark with polarization $\sigma$ in the
background quark or gluon field of the target is written as
\begin{align} 
  \label{e:Vsigma}
  V_{\ul x} (\sigma) \equiv V_{\ul x} + \sigma V_{\ul x}^{pol}
\end{align}
with the polarization-dependent part of that background-field
propagator being more than a pure gauge rotation.  The polarized
Wilson line may couple once to a transverse component $A_\bot^i$ of
the gluon field, or it may exchange two $t$-channel quarks with the
target.  Since the leading high-energy behavior of the quark
propagator is spin-independent, $V_{\ul x}^{pol}$ is suppressed
relative to $V_{\ul x}$ by a factor of the quark energy.  Each
additional spin-dependent interaction is further suppressed by a power
of the quark energy, so $V_{\ul x}^{pol}$ contains exactly one
spin-dependent interaction (a gluon exchange or a two-quark exchange),
along with any number of eikonal, spin-independent gluon-exchange
interactions.  The double angle brackets in \eqref{e:Gdef1} are
defined to remove this suppression of $1/z s$ from the dipole trace
\cite{Kovchegov:2015pbl},
\begin{align}
  \llangle \mathcal{O} \rrangle (z) \equiv z s \: \Big\langle
  \mathcal{O} \Big\rangle (z) ,
\end{align}
while the single angle brackets $\langle \ldots \rangle$ denote the
averaging in the (polarized) target proton or nucleus. Note that $z$
used in the rescaling is the momentum fraction of the polarized line
in the dipole, while $z$ in the argument is the smallest momentum
fraction between the polarized and unpolarized lines
\cite{Kovchegov:2015pbl}: these two $z$ values could be different.

To further simplify Eqs.~\eqref{e:observables} it would be tempting to
replace
\begin{align}
  \label{eq:repl1}
  \llangle \tr \left[ V_{\ul 0} V_{\ul 1}^{pol \, \dagger} \right]
  \rrangle^* (z) \: \to \: \llangle \tr \left[ V_{\ul 1}^{pol} V_{\ul
      0}^\dagger \right] \rrangle (z)
\end{align}
as is often done for the unpolarized small-$x$ evolution (the asterisk
denotes complex conjugation). However, here one has to be more
careful: for a general target state $|T\rangle$ we have
\begin{align}
  \label{eq:repl2}
  \langle T | \tr \left[ V_{\ul 0} V_{\ul 1}^{pol \, \dagger} \right]
  |T \rangle^* = \langle {\bar T} | \tr \left[ V_{\ul 1}^{pol} V_{\ul
      0}^\dagger \right] |{\bar T} \rangle ,
\end{align}
where $|\bar T \rangle$ denotes the charge-conjugate target
state. While unpolarized BFKL, BK and JIMWLK evolution is insensitive
to whether the target is, say, a quark or an anti-quark, this is not
the case for helicity evolution. For instance, the $t$-channel quark
exchange shown in \fig{f:helicity_TMD} is possible for the quark
target but is impossible for the anti-quark one.

Keeping this in mind, let us consider the flavor singlet case, 
\begin{align}
  \Delta q^{S} (x, Q^2) \equiv \sum_f \left[ \Delta q^f (x, Q^2) +
    \Delta \bar{q}^f (x, Q^2) \right] .
\end{align}
Adding to diagrams in \fig{f:helicity_TMD} the graphs which have the
quark loop particle number flow in the opposite direction (such that
the tagged particle is an anti-quark) and summing over all flavors
simplifies Eqs.~\eqref{e:observables} to
\begin{subequations} \label{e:observables_S}
  \begin{align} 
    g^S_1 (x, Q^2) &= \frac{N_c}{2 \, \pi^2 \alpha_{EM}} \sum_f
    \int\limits_{z_i}^1\frac{dz }{z^2 (1-z)} \, \int d x_{01}^2 \,
    \left[ \half \sum_{\lambda \sigma \sigma'} | \psi_{\lambda \sigma
        \sigma'}^T |^2_{(x_{01}^2 , z)} + \sum_{\sigma \sigma'}
      |\psi_{\sigma \sigma'}^L|^2_{(x_{01}^2 , z)} \right] G(x_{01}^2
    , z) ,
  \\
  \Delta q^S (x, Q^2) &= \frac{N_c}{2 \pi^3} \sum_f
  \int\limits_{z_i}^1 \frac{dz}{z} \int\limits_{\tfrac{1}{z
      s}}^{\tfrac{1}{z Q^2}} \frac{d x_{01}^2}{x_{01}^2} \, G(x_{01}^2
  , z) ,
  \\
  g_{1L}^S (x, k_T^2) &= \frac{8 \, N_c}{(2\pi)^6} \sum_f
  \int\limits_{z_i}^1 \frac{dz}{z} \int d^2 x_{01} \, d^2 x_{0'1} \:
  e^{-i \ul{k} \cdot (\ul{x}_{01} - \ul{x}_{0' 1})} \:
  \frac{\ul{x}_{01} \cdot \ul{x}_{0'1}}{x_{01}^2 x_{0'1}^2} \,
  G(x_{01}^2 , z) . \label{g1LS}
  \end{align}
\end{subequations}

We see from Eqs.~\eqref{e:observables_S} that the small-$x$ polarized
scattering dynamics are contained within the {\sl polarized dipole
  amplitude}, which is defined by
\begin{subequations}\label{e:Gdef}
  \begin{align} 
    \label{e:Gdef1} 
    & G_{10} (z) \equiv \frac{1}{2 N_c} \llangle \tr \left[ V_{\ul 0}
      V_{\ul 1}^{pol \, \dagger} \right] + \tr \left[V_{\ul 1}^{pol}
      V_{\ul 0}^\dagger \right] \rrangle (z) = G(\ul{x}_1 , \ul{x}_0 ,
    z) = G(\ul{x}_{10} , \ul{b} , z) ,
   \\ \label{e:Gdef2}
   & G (x_{01}^2 , z) \equiv \int d^2 b \: G_{10} (z) ,
 \end{align}
\end{subequations}
where, again, $\ul{b} = \thalf (\ul{x}_1 + \ul{x}_0)$ is the impact
parameter of the dipole which is held fixed in $G_{10} (z)$ and
integrated in $G(x_{01}^2 , z)$. In arriving at Eqs.~\eqref{e:Gdef} we
have assumed that $G_{10} (z)$ and, hence, $G(x_{01}^2 , z)$ are both
real, $G_{10} (z) = G^*_{10} (z)$, which is true for the leading
contributions to $G_{10} (z)$ without evolution (the initial
conditions) and is still the case after evolving the polarized dipole
amplitude using helicity evolution \cite{Kovchegov:2015pbl}.
The helicity evolution equations constructed in
\cite{Kovchegov:2015pbl} concentrated on the flavor singlet case of
Eqs.~\eqref{e:observables_S} and \eqref{e:Gdef}.

Although the polarized Wilson line defining $G_{10}(z)$ is difficult
to define operatorially, it corresponds to the spin-dependent part of
 the $S$-matrix of a quark propagating through the background field of 
the target. Therefore, we can define
$G_{10}(z)$ indirectly by relating it to the dipole cross-section via
the optical theorem:
\begin{align}
  \frac{1}{N_c} \Big\langle \tr \left[ V_{\ul 0} (\sigma_0) \, V_{\ul
      1}^\dagger (\sigma_1) \right] \Big\rangle (z) \, &\equiv \, S
  \Big[ q_{\ul{0}} (\sigma_0) , \bar{q}_{\ul{1}} (\sigma_1) , z s
  \Big]
  %
\approx
  1 - \mathrm{Im} \, T \Big[ q_{\ul{0}} (\sigma_0) , \bar{q}_{\ul{1}}
  (\sigma_1) , z s \Big]
  \notag \\ & =
  1 - \half \frac{d \sigma}{d^2 b} \Big[ q_{\ul{0}} (\sigma_0) ,
  \bar{q}_{\ul{1}} (\sigma_1) , z s \Big] , \label{conn2sigma}
\end{align}
where we have neglected the real part of the (expectation value of
the) $T$-matrix at high energies as being higher-order in the strong
coupling $\as$.\footnote{There is a subtlety here: the real part of the
  unpolarized $T$-matrix for eikonal Wilson lines, the odderon
  \cite{Hatta:2005as,Kovchegov:2003dm}, is $\as$-suppressed compared
  to the leading unpolarized imaginary part retained in
  \eq{conn2sigma}. In our power counting, this makes the unpolarized
  real part much larger than the leading polarization-dependent
  imaginary part we are interested in, since the latter is
  energy-suppressed. Hence, the approximation in \eq{conn2sigma}
  should be understood as correctly retaining only the leading
  polarized and unpolarized contributions.} Here $q_{\ul x} (\sigma)$
($\bar q_{\ul x} (\sigma)$) denotes a quark (antiquark) at transverse
position $\ul{x}$ and spin $\sigma$. Using \eqref{e:Vsigma} to expand
the Wilson lines on the left-hand side, we obtain
\begin{align}
  \frac{1}{N_c} \Big\langle \tr \left[ V_{\ul 0} \, V_{\ul 1}^{pol
      \dagger} \right] \Big\rangle (z) \, = - \frac{1}{4} \,
  \sum_{\sigma_0 \sigma_1} \, \sigma_1 \, \frac{d \sigma}{d^2 b} \Big[
  q_{\ul{0}} (\sigma_0) , \bar{q}_{\ul{1}} (\sigma_1) , z s \Big]
  %
  \equiv
  - \frac{d \sigma}{d^2 b} \Big[ q_{\ul{0}}^{unp} ,
  \Delta\bar{q}_{\ul{1}} , z s \Big] , \label{VVsigma}
\end{align}
and similarly for $\frac{1}{N_c} \Big\langle \tr \left[ V_{\ul
    1}^{pol} \, V_{\ul 0}^{\dagger} \right] \Big\rangle (z)$.  This
gives an expression for the polarized dipole amplitude in terms of the
spin-dependent part of the dipole cross-section:
\begin{align} 
  \label{e:GdefCS}
  G_{10} (z) = - \frac{z s}{2} \left( \frac{d \sigma}{d^2 b} \Big[
    q_{\ul{0}}^{unp} , \Delta\bar{q}_{\ul{1}} , z s \Big] + \frac{d
      \sigma}{d^2 b} \Big[ {\bar q}_{\ul{0}}^{unp} , \Delta q_{\ul{1}}
    , z s \Big] \right).
\end{align}
%

\begin{figure}[tb]
\centering
\includegraphics[width= 0.9 \textwidth]{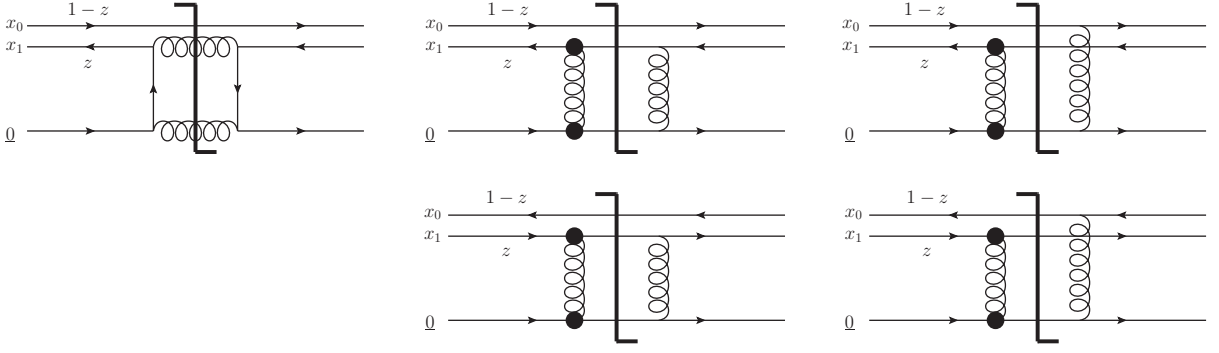}
\caption{Diagrams contributing to the lowest-order initial conditions
  $G_{01}^{(0)} (z)$ for a quark target.  The top line of diagrams
  contributes to $\tr[V_0 V_1^{pol \, \dagger}]$, and the bottom line
  contributes to $\tr[V_1^{pol} V_0^{\dagger}]$.  The black circles
  denote non-eikonal quark-gluon vertices which transfer spin, and
  complex-conjugates must be added to the asymmetric diagrams.  Note
  that in the second line, the quark and antiquark lines have been
  interchanged, consistent with the definition \eqref{e:Gdef1}.}
\label{f:init_conds}
\end{figure}

With the help of \eqref{e:GdefCS}, we can calculate the polarized
dipole amplitude at lowest order for a quark target as shown in
Fig.~\ref{f:init_conds}. For simplicity the target quark is assumed to
be at the origin in the transverse plane. These explicit expressions
for $G_{10}^{(0)}$ can serve as initial conditions for the subsequent
small-$x$ evolution:
\begin{subequations} \label{e:IC}
  \begin{align} \label{e:IC1} G_{10}^{(0)} &= \frac{\alpha_s^2
      C_F}{N_c} \left[\frac{C_F}{x_1^2} - 2 \, \pi \, \delta^2
      (\ul{x}_1) \, \ln (z s \, x_{10}^2) \right]
  \\ \label{e:IC2}
  G^{(0)} (x_{10}^2 , z) &= \frac{\alpha_s^2 C_F}{N_c} \pi \left[ C_F
    \ln\frac{z s}{\Lambda^2} - 2 \, \ln (z s \, x_{10}^2) \right] ,
 \end{align}
\end{subequations}
where the impact parameter integral $\int d^2 b = \int d^2 x_0 = \int
d^2 x_1$ is cut off in the UV by the energy $b^2 > \tfrac{1}{z s}$ and
in the IR by a cutoff $b^2 < \tfrac{1}{\Lambda^2}$, while $C_F =
(N_c^2 -1)/2 N_c$.  One can also dress these linearized expressions
with quasi-classical multiple Glauber-Mueller (GM) rescatterings
\cite{Mueller:1989st} in the spirit of the McLerran-Venugopalan (MV)
model
\cite{McLerran:1993ni,McLerran:1993ka,McLerran:1994vd,Kovchegov:1996ty,Kovchegov:1997pc,Jalilian-Marian:1997xn},
obtaining
\begin{align}
  G_{10}^{(0)} &= \frac{\alpha_s^2 C_F}{N_c} \left[ \frac{C_F}{x_1^2}
    - 2 \, \pi \, \delta^2 (\ul{x}_1) \, \ln (z s \, x_{10}^2) \right]
  \, \exp\left[-\tfrac{1}{4} \, x_{10}^2 \, Q_s^2 (b) \,
    \ln\tfrac{1}{x_{01} \Lambda} \right] \label{e:IC3}
\end{align}
where $Q_s$ is the saturation scale before evolution. \eq{e:IC3}
includes saturation effects by resumming multiple rescatterings and
can also serve as the initial condition for small-$x$ helicity
evolution, if the latter includes saturation effects as well.

\subsection{Flavor Singlet Helicity Evolution at Small $x$}
\label{sec:FSevol}

\begin{figure}[bt]
 \centering
 \includegraphics[width=0.8\textwidth]{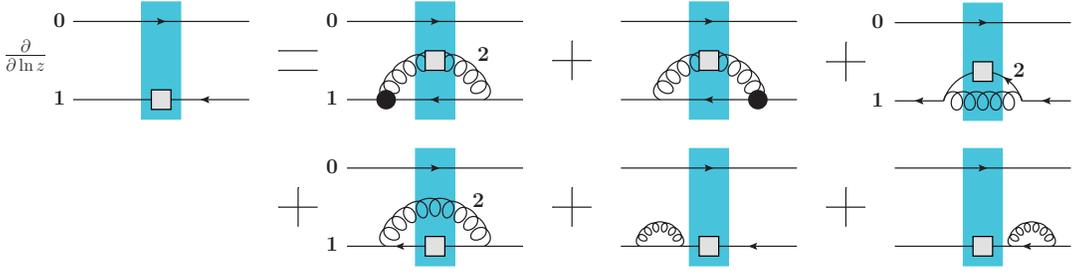}
 \caption{One step of small-$x$ evolution in the polarized dipole
   amplitude $G_{10}$.  The thick vertical rectangle represents the
   shock wave interaction with the target, the large black circle
   vertices represent the sub-eikonal emission of a polarized gluon,
   and the small gray box denotes the polarized Wilson line.  For
   simplicity, the initial condition $G_{10}^{(0)}$ is not shown.}
\label{f:evol_step_latex}
\end{figure}

As derived in \cite{Kovchegov:2015pbl}, the small-$x$ evolution of the
polarized dipole amplitude resums double logarithms of the energy:
$\alpha_s \ln^2 \tfrac{s}{\Lambda^2} \sim \alpha_s \ln^2\tfrac{1}{x}$.
The polarized evolution proceeds by the radiation of longitudinally
soft polarized partons with momentum fractions $z' \ll z$ (top line of
Fig.~\ref{f:evol_step_latex}); there are also non-vanishing
double-logarithmic contributions from the radiation of longitudinally
soft {\it unpolarized} gluons akin to the unpolarized BFKL / BK /
JIMWLK equations (bottom line of Fig.~\ref{f:evol_step_latex}). The
contribution of other polarized and unpolarized gluon emission
diagrams amounts to introducing an IR cutoff $x_{21} < x_{10}$ on the
$x_{21}$-integral in the gluon-emission diagrams in
Fig.~\ref{f:evol_step_latex} \cite{Kovchegov:2015pbl}: for brevity we
do not show those remaining graphs.  The result of one step of
double-logarithmic (DLA) evolution in the polarized dipole amplitude
is given by \cite{Kovchegov:2015pbl}
\begin{align} 
  \label{e:opevol1}
  & G_{10} (z) = G_{10}^{(0)} (z) + \frac{\alpha_s}{2 \pi^2}
  \int\limits_{\Lambda^2 /s}^z \frac{d z'}{z'} \, \int\frac{d^2
    x_{2}}{x_{21}^2} \: \theta\left( x_{21}^2 - \tfrac{1}{z' s}
  \right)
      \notag \\ & \:\:\: \times \bigg\{
      \theta (x_{10} - x_{21}) \, \frac{1}{N_c} \llangle \tr \left[
        t^b \, V_{\ul{0}} \, t^a \, V_{\ul{1}}^{\dagger} \right]
      \left( U^{pol}_{\ul{2}} \right)^{ba} + \tr \left[ t^b \,
        V_{\ul{1}} \, t^a \, V_{\ul{0}}^{\dagger} \right] \left(
        U^{pol \, \dagger}_{\ul{2}} \right)^{ab} \rrangle (z')
      \notag \\ & \:\:\: +
      \theta (x_{10}^2 z - x_{21}^2 z') \, \frac{1}{4 N_c} \bigg[
      \llangle \tr\left[ V_{\ul 0} V_{\ul 1}^\dagger \right]\tr\left[
        V_{\ul 1} V_{\ul 2}^{pol \, \dagger} \right] + \tr\left[
        V_{\ul 1} V_{\ul 0}^\dagger \right]\tr\left[ V_{\ul 2}^{pol}
        V_{\ul 1}^\dagger \right] \rrangle (z')
      \notag \\ & \hspace{4.2cm} -
      \frac{1}{2N_c} \llangle \tr\left[V_{\ul 0} V_{\ul 2}^{pol \,
          \dagger}\right] + \tr\left[V_{\ul 2}^{pol} V_{\ul
          0}^{\dagger}\right] \rrangle (z') \bigg]
      \notag \\ & \:\:\: +
      \theta (x_{10} - x_{21}) \frac{1}{N_c} \bigg[ \llangle \tr
      \left[ V_{\ul{0}} \, V^{\dagger}_{\ul{2}} \right] \, \tr \left[
        V_{\ul{2}} \, V^{pol \, \dagger}_{\ul{1}} \right] + \tr \left[
        V_{\ul{2}} \, V^{\dagger}_{\ul{0}} \right] \, \tr \left[
        V^{pol}_{\ul{1}} \, V^{\dagger}_{\ul{2}} \right] \rrangle (z')
      \notag \\ & \hspace{4.2cm} -
      N_c \llangle \tr \left[ V_{\ul{0}} \, V_{\ul{1}}^{pol \,
          \dagger} \right] + \tr \left[ V^{pol}_{\ul{1}} \,
        V_{\ul{0}}^{\dagger} \right] \rrangle (z') \bigg] \: \bigg\}
\end{align}
as drawn diagrammatically in Fig.~\ref{f:evol_step_latex}.  The
polarized adjoint Wilson line $U_{\ul 2}^{pol}$ is defined analogously
to \eqref{e:Vsigma}.  Like equations in the Balitsky hierarchy
\cite{Balitsky:1996ub,Balitsky:1998ya} for unpolarized small-$x$
evolution, the evolution of the polarized dipole $G_{10}$ is not
closed, coupling to increasingly complex operators at each step of
evolution.  The first term in braces in \eqref{e:opevol1} corresponds
to the radiation of a soft polarized gluon, as shown in the first two
diagrams of Fig.~\ref{f:evol_step_latex}.  The second term corresponds
to the radiation of a soft polarized (anti)quark, as shown in the
third diagram of Fig.~\ref{f:evol_step_latex}.  The last term in
braces in \eqref{e:opevol1} corresponds to the radiation of soft
unpolarized gluons, as shown in the second row of diagrams in
Fig.~\ref{f:evol_step_latex}.  As we have already mentioned, the
diagrams in the first and third classes are DLA in the $x_{21} <
x_{10}$ portion of the full phase space $x_{10}^2 z \gg x_{21}^2 z'$
due to partial cancellations from other diagrams which we do not show
explicitly.

Equation (\ref{e:opevol1}) does not close, and represents the
lowest-order equation in the infinite tower of equations involving
higher and higher order operators (in the number of Wilson lines), the
helicity evolution analogue of the unpolarized Balitsky hierarchy
\cite{Balitsky:1996ub}. This helicity hierarchy, represented here by
the evolution equation \eqref{e:opevol1}, is difficult to solve; at
the moment it is too early to tell whether it would be suitable for
stochastic methods used in solving the unpolarized JIMWLK evolution
\cite{Weigert:2000gi}.  It can, however, be solved in limits in which
the operator hierarchy closes: namely the large-$N_c$ limit and the
large-$N_c \,\&\,N_f$ limit \cite{Kovchegov:2015pbl}.  In the
large-$N_c$ limit the evolution is gluon-driven
(cf. \cite{Mueller:1994rr,Mueller:1994jq,Mueller:1995gb}). Assuming
that the parent dipole $10$ in \fig{f:evol_step_latex} comes from the
quark and anti-quark lines of different gluons at large $N_c$, we can
neglect the radiation of soft quarks (second term in braces in
\eqref{e:opevol1}) and simplify the remaining terms, obtaining
\cite{Kovchegov:2015pbl}
\begin{subequations} \label{e:largeNevol}
  \begin{align} \label{e:largeNevol1} G_{10} (z) &= G_{10}^{(0)} (z) +
    \frac{\alpha_s N_c}{2\pi^2} \int\limits_{\Lambda^2/s}^z \frac{d
      z'}{z'} \int\frac{d^2 x_2}{x_{21}^2} \, \theta(x_{10} - x_{21})
    \, \theta(x_{21}^2 - \tfrac{1}{z' s})
   \notag \\ &\times
   [ 2 \Gamma_{20 , \, 21} (z') S_{21} (z') + 2 G_{21} (z') S_{02}
   (z') + G_{12}(z') S_{02} (z') - \Gamma_{10, \, 21}(z') ]
   \\ \label{e:largeNevol2}
   \Gamma_{20 , \, 21} (z') &= G_{20}^{(0)} (z') + \frac{\alpha_s
     N_c}{2\pi^2} \int\limits_{\Lambda^2/s}^{z'} \frac{d z''}{z''}
   \int\frac{d^2 x_3}{x_{32}^2} \, \theta\left( \min[x_{02}^2 ,
     x_{21}^2 \tfrac{z'}{z''}] - x_{32}^2 \right) \, \theta(x_{32}^2 -
   \tfrac{1}{z'' s})
   \notag \\ &\times
   [ 2 \Gamma_{30 , \, 32} (z'') S_{23} (z'') + 2 G_{32} (z'') S_{03}
   (z'') + G_{23} (z'') S_{03} (z'') - \Gamma_{20, \, 32} (z'') ] ,
 \end{align}
\end{subequations}
where the unpolarized dipole scattering amplitude
\begin{align} \label{e:MV} S_{21} (z) &\equiv \frac{1}{N_c}
  \left\langle \tr[V_{\ul 2} V_{\ul 1}^\dagger] \right\rangle \approx
  S_{12} (z)
\end{align}
is obtained from the BK/JIMWLK evolution equations
\cite{Balitsky:1996ub,Balitsky:1998ya,Kovchegov:1999yj,Kovchegov:1999ua,Jalilian-Marian:1997dw,Jalilian-Marian:1997gr,Iancu:2001ad,Iancu:2000hn}
with the initial condition given by the GM/MV result
\cite{Mueller:1989st,McLerran:1998nk}
\begin{align}
  S^{(0)}_{10} (z) = \exp\left[ -\tfrac{1}{4} x_{10}^2 Q_s^2 (\ul{b})
    \, \ln\tfrac{1}{x_{10} \Lambda} \right]
\end{align}
which is independent of $z$. Since BK and JIMWLK evolution is
leading-logarithmic (LLA) at the leading order, it does not contribute
in the strict DLA limit, in which we simply put $S=1$ in
Eqs.~\eqref{e:largeNevol}. However, when the precision of helicity
evolution is increased beyond DLA to LLA level, saturation effects
would come in through $S_{21} (z)$ as shown in
Eqs.~\eqref{e:largeNevol}.

Even in the large-$N_c$ limit, the operator evolution
\eqref{e:opevol1} remains a system of equations because the dipoles
are not all independent of each other.  The general phase space which
yields DLA contributions
\begin{align} 
  \label{e:ordering}
  z_1 x_{1T}^2 \gg z_2 x_{2T}^2 \gg z_3 x_{3T}^2 \cdots \hspace{2cm} 1
  \gg z_1 \gg z_2 \gg z_3 \cdots
\end{align}
competes with the $\theta (x_{10} - x_{21})$ functions of
\eqref{e:opevol1} which arise from partial cancellations with other
diagrams.  In some cases, the ordering \eqref{e:ordering} is more
restrictive than the $\theta$ functions, which introduces an extra
dependence of one dipole amplitude on the dipole size of another.
This leads to the ``neighbor dipole'' function $\Gamma_{20, \, 21}
(z')$ \cite{Kovchegov:2015pbl}.  (Note: the labeling here is different
than in \cite{Kovchegov:2015pbl}.  Here the first index of $\Gamma$
denotes the polarized line, bringing it into consistency with $G_{10}$
defined in \eqref{e:Gdef1}.)  In this term, further evolution
continues in the large dipole $x^2_{20}$, but residual dependence on
the size of the neighbor dipole $x_{21}^2$ remains through the limits
of integration in \eqref{e:largeNevol2}.  Thus even in the large-$N_c$
limit, helicity evolution is, in this respect, more complex than
unpolarized evolution.  We note that the virtual corrections (last
term in brackets) can be shown to enter as neighbor dipole functions
(see Sec.~\ref{sec:Cancellations}).

There is another interesting feature in the
Eqs.~\eqref{e:largeNevol}. Let us first note that the small-$x$
polarized leading-order DGLAP splitting functions for gluon emission
are $\Delta P_{Gq} (z\to 0) = 2 C_F \, (\as/2 \pi)$ and $\Delta P_{GG}
(z\to 0) = 4 N_c \, (\as /2 \pi)$ \cite{Altarelli:1977zs}. Hence, in
the large-$N_c$ limit we have $\Delta P_{GG} (z\to 0) = 4 \, \Delta
P_{Gq} (z\to 0)$. The difference between the two splitting functions
is not simply due to the difference of their color factors, $C_F
\approx N_c/2$ and $N_c$ respectively, as is the case for the
unpolarized small-$x$ splitting functions (for which $P_{GG} (z\to 0)
= 2 \, P_{Gq} (z\to 0)$): this would only account for a factor of 2
difference. The other factor of 2 comes from the helicity dynamics of
the $G \to GG$ splitting as compared to the $q \to Gq$
splitting. Thus, for the large-$N_c$ limit, which is dominated by
gluon dynamics, it is insufficient to simply take \eq{e:opevol1} and
send $N_c \to \infty$. There is an essential difference between the
evolution of the polarized quark dipole and the polarized dipole made
out of ``quark lines'' in the large-$N_c$ gluon dipole: the splitting
in the latter come with an extra factor of 2. This is the reason for
the extra factor of 2 in front of the first two terms in the
integrands of both Eqs.~\eqref{e:largeNevol}. In the large-$N_c$ limit
one should understand the polarized dipole definition \eqref{e:Gdef1}
as involving eikonal quark and anti-quark lines coming from gluon
lines. To derive this factor of 2 more formally one needs to start
with the analogue of \eq{e:opevol1} for the polarized gluon dipole and
take the large-$N_c$ limit: this is presented in
Appendix~\ref{sec:largeN}.


\subsection{Solution of flavor singlet helicity evolution equations at
  large $N_c$}
  
\label{sec:FSsol}

To facilitate solving the large-$N_c$ equations \eqref{e:largeNevol},
let us first linearize them by dropping the unpolarized multiple
rescattering terms like $S_{21}$ and then integrate over the impact
parameter $\ul{b}$. This is justified outside of the saturation
region, where $S \approx 1$. Note also that the $S$-terms are LLA and
should be put to one in the strict DLA limit. Doing so, we obtain
\begin{subequations} \label{e:linear}
  \begin{align} 
    \label{e:linear1}
    G(x_{10}^2 , z) &= G ^{(0)} (x_{10}^2 , z) + \frac{\alpha_s
      N_c}{2\pi} \int\limits_{\tfrac{1}{x_{10}^2 s}}^{z} \frac{d
      z'}{z'} \int\limits_{\tfrac{1}{z' s}}^{x_{10}^2} \frac{d
      x_{21}^2}{x_{21}^2} \left[ \Gamma (x_{10}^2 , x_{21}^2 , z') + 3
      G(x_{21}^2 , z') \right] ,
   \\ \label{e:linear2}
   \Gamma (x_{10}^2 , x_{21}^2 , z') &= G^{(0)} (x_{10}^2, z') +
   \frac{\alpha_s N_c}{2\pi} \int\limits_{\tfrac{1}{x_{10}^2 s}}^{z'}
   \frac{d z''}{z''} \hspace{-0.5cm} \int\limits_{\tfrac{1}{z''
       s}}^{\min\left[x_{10}^2 \, , \, x_{21}^2
       \tfrac{z'}{z''}\right]} \hspace{-0.9cm} \frac{d
     x_{32}^2}{x_{32}^2} \left[ \Gamma (x_{10}^2 , x_{32}^2 , z'') + 3
     G(x_{32}^2 , z'') \right] ,
 \end{align}
\end{subequations}
where we have neglected the small differences between the large dipole
sizes $x_{01}^2 \approx x_{02}^2 \approx x_{03}^2$.  

The usual Laplace-Mellin transform technique fails to simplify the
system \eqref{e:linear} due to the presence of the neighbor dipole
function $\Gamma (x_{10}^2 , x_{21}^2 , z')$.  Instead we resort to
solving \eqref{e:linear} numerically by discretizing the independent
variables on a lattice.  Since the $z, z'$ dependence enters through
the upper limits of the $z' , z''$ integrations, respectively, these
equations are well-suited to solution by iteration: starting with just
the initial conditions at $z = \tfrac{1}{x_{10}^2 s}$, we can
systematically compute the polarized dipole amplitude at $z$ using the
already-tabulated results for lower values of $z$.  By evolving to
sufficiently large $z s$, we look for the emergence of power-law
behavior $G(x_T^2 , z s) \propto (z s)^{\alpha_{h}}$ and extract the
helicity intercept $\alpha_{h}$ by performing a linear fit to $\ln G$.
For further details of the numerics and for the implications regarding
the quark contribution $S_q$ to the proton spin, we refer the
interested reader to the accompanying Letter \cite{Letter}.

The high-energy asymptotics of the polarized dipole amplitude found in
\cite{Letter} can be summarized by
\begin{align}
  \label{hel_int}
  G (x_{10}^2 , z) \propto ( z s )^{\alpha_{h}} \hspace{1cm}
  \mathrm{with} \hspace{1cm} \alpha_{h} \approx 2.31 \:
  {\sqrt{\tfrac{\alpha_s N_c}{2 \pi}}} .
\end{align}
Using Eqs.~\eqref{e:observables_S} we conclude that the small-$x$
asymptotics of flavor-singlet helicity observables is
\begin{align}
  g_1^S (x, Q^2) \sim \Delta q^S (x, Q^2) \sim g_{1L}^S (x, k_T^2)
  \sim \left( \frac{1}{x} \right)^{\!\alpha_{h}} \approx \left(
    \frac{1}{x} \right)^{\!2.31 \: {\sqrt{\tfrac{\alpha_s N_c}{2
          \pi}}}}.
\end{align}
This is one of the main results of this project so far.

Our evolution equations \eqref{e:opevol1} only close in the
large-$N_c$ or large-$N_c \& N_f$ limits. In \cite{Letter} we have
only solved them numerically in the large-$N_c$ (pure glue) limit
obtaining the result given in \eq{hel_int}. Solution of the evolution
equations derived in \cite{Kovchegov:2015pbl} for the large-$N_c \&
N_f$ limit is left for future work.

The intercept in \eq{hel_int} is smaller by about $35\%$ than the pure
glue intercept obtained by BER in \cite{Bartels:1996wc},
$\alpha_{h}^{BER} = 3.66 \sqrt{\tfrac{\alpha_s N_c}{2\pi}}$. Despite
this disagreement on the full result,\footnote{There is a caveat here:
  our result \eqref{hel_int} for the intercept was calculated in the
  large-$N_c$ pure-glue limit; the part of the calculation in
  \cite{Bartels:1996wc} leading to the intercept $\alpha_{h}^{BER} =
  3.66 \sqrt{\tfrac{\alpha_s N_c}{2\pi}}$ was for the pure glue case,
  but was not in the large-$N_c$ limit. Therefore one could attribute
  the difference between the two numbers to the difference between the
  large-$N_c$ limit (us) and $N_c =3$ (BER). To explore this
  possibility we have reproduced BER's solution for pure glue
  obtaining 
  \begin{align}
    \label{eq:alpha_BER}
    \alpha_{h}^{BER} = \sqrt{\tfrac{17 + \sqrt{97}}{2}}
    \sqrt{\tfrac{\alpha_s N_c}{2\pi}} \approx 3.66
    \sqrt{\tfrac{\alpha_s N_c}{2\pi}}.
  \end{align}
  Hence 3.66 is a pure number and this result holds for any $N_c$ in
  the BER framework. Therefore, disagreement between the BER intercept
  and ours in \eq{hel_int} is not due to the large-$N_c$ limit
  employed in our case.} we agree with BER on important subsets of the
calculation such as the ``ladder graphs'' which include DGLAP-like
quark/gluon mixing \cite{Kovchegov:2015pbl} and the flavor non-singlet
helicity evolution intercept which we obtain below in \eq{e:Reggeon}.

Given this discrepancy, it is important to validate the internal
consistency of our calculation and to compare with the results of BER
wherever possible.  Direct comparison is difficult on a term-by-term
or diagram-by-diagram basis, since we work in different gauges (the
light-cone gauge versus Feynman gauge) and use very different
formalisms ($s$-channel light-front wave functions versus infrared
evolution equations).  There are, however, some consistency checks we
can do to increase the confidence in our result and to better
understand the nature of our evolution equations.  We will pursue
these cross-checks next in the following sub-sections, where we will
justify the neighbor dipole amplitude following virtual correction in
the evolution of \fig{f:evol_step_latex} and successfully re-derive
the small-$x$ polarized DGLAP anomalous dimension $\Delta P_{GG} (z
\to 0)$ at NLO. Further comparison with the calculation by BER can be
found in Appendix~\ref{sec:BERcheck}.


\subsection{Cross-Check: Virtual Diagrams and Real-Virtual
  Cancellations}
\label{sec:Cancellations}

\subsubsection{Evolution Subsequent to a Virtual Correction}

First we would like to cross-check and clarify the origin of the
neighbor dipole amplitude $\Gamma$ in the last term of the integrand
in both Eqs.~\eqref{e:largeNevol}.  These terms arise from the
evolution subsequent to the virtual corrections in the last two
diagrams of the bottom line in \fig{f:evol_step_latex}.  The real
correction (left-most diagram in the bottom line of
\fig{f:evol_step_latex}) imposes a lifetime ordering constraint
\eqref{e:ordering} on further DLA evolution of the dipole 21, while
naively it seems that the virtual corrections (two right-most diagrams
in the bottom line of \fig{f:evol_step_latex}) impose no such
constraint.  As we shall see, the virtual diagrams actually {\it{do}}
impose the same lifetime ordering condition in order for the
subsequent evolution to remain DLA: to see this we need to perform a
calculation.

\begin{figure}[hbt]
 \centering
 \includegraphics[width=\textwidth]{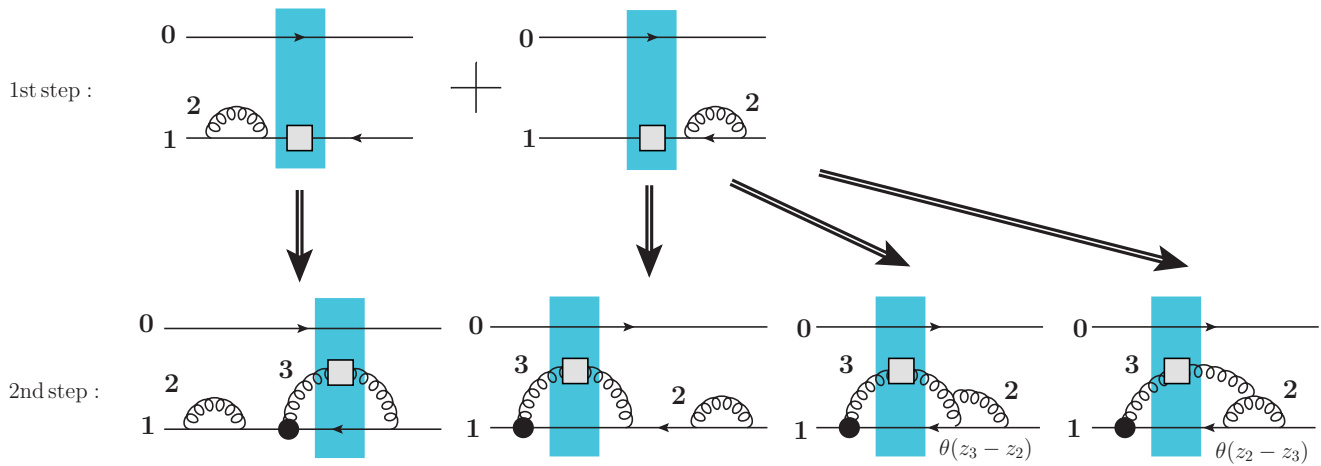}
 \caption{Two steps of helicity evolution.  In the first step, we
   consider BFKL-like virtual corrections (last 2 diagrams in
   Fig.~\ref{f:evol_step_latex}).  In the second step, we consider one
   particular type of correction, the emission of a soft polarized
   gluon (first diagram in Fig.~\ref{f:evol_step_latex}), connected in
   all possible ways.  The last two diagrams differ only in the
   time-ordering of the vertices.}
\label{f:real_virt_cancel}
\end{figure}

Consider one particular step of subsequent DLA evolution following the
virtual corrections, as shown in Fig.~\ref{f:real_virt_cancel}: the
emission of a soft polarized gluon.  One may have a ladder-type
correction, as shown in the first two diagrams of the second line of
Fig.~\ref{f:real_virt_cancel}.  Or one may have a non-ladder-type
correction with the polarized gluon attaching to the virtual gluon, as
shown in the last two diagrams.  Let us choose $z_2 \gg z_3$ for
specificity (one step of evolution) and compute these diagrams
explicitly in LFPT to see exactly what the DLA regime of the second
evolution step is.  We will work in the large-$N_c$ limit, which is
the context in which our evolution equations \eqref{e:largeNevol} and
\eqref{e:linear} are derived.

\begin{figure}[hbt]
 \centering
 \includegraphics[width=\textwidth]{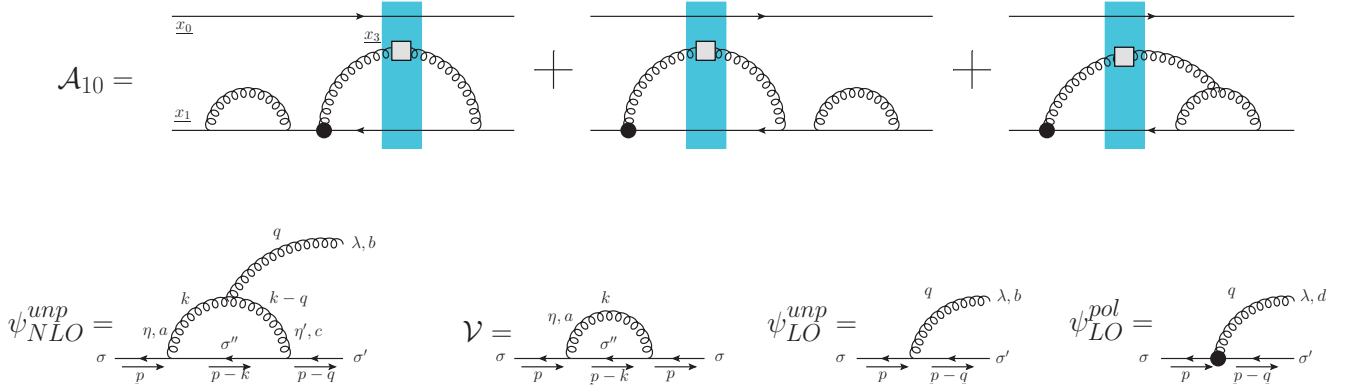}
 \caption{Calculation of the light-front wave functions which go into
   the diagrams containing a virtual correction and a subsequent
   evolution step.}
\label{f:real_virt_WF}
\end{figure}

The polarized dipole amplitude $\mathcal{A}_{1 0}$ generated by the
three diagrams shown in Fig.~\ref{f:real_virt_WF} is given by
\begin{align} \label{e:realvirt1} \mathcal{A}_{1 0} &= \int
  \!\!\frac{d q^+}{4 \pi q^+} d^2 x_3 \sum_{\mathrm{colors}}
  \sum_{\lambda \sigma'} \left\langle U_{\ul 3} (\lambda) \,
    V_1^\dagger (\sigma') \right\rangle (\tfrac{q^+}{p^+}) \: \bigg[
  \mathcal{V} \times \tilde\psi_{LO}^{pol} (\ul{x}_{31},
  \tfrac{q^+}{p^+}) \times \left( - \tilde\psi_{LO}^{unp}
    (\ul{x}_{31}, \tfrac{q^+}{p^+} ) \right)^*
  \notag \\ & +
  \tilde\psi_{LO}^{pol} (\ul{x}_{31} , \tfrac{q^+}{p^+}) \times \left(
    - \tilde\psi_{LO}^{unp} (\ul{x}_{31} , \tfrac{q^+}{p^+} )
  \right)^* \times \mathcal{V}^* + \tilde\psi_{LO}^{pol} (\ul{x}_{31},
  \tfrac{q^+}{p^+}) \times \left( - \tilde\psi_{NLO}^{unp}
    (\ul{x}_{31} , \tfrac{q^+}{p^+}) \right)^* \bigg] ,
\end{align}
where $z_3 = \tfrac{q^+}{p^+}$ is the momentum fraction of the
polarized gluon, and the notation is otherwise indicated in
Fig.~\ref{f:real_virt_WF}.  The coordinate-space wave functions are
related to the momentum-space wave functions by
\begin{align}
  \tilde\psi (\ul{x}_{31}, z_3) = \int\!\!\frac{d^2 q}{(2\pi)^2} \,
  e^{i \ul{q} \cdot \ul{x}_{31}} \, \psi (\ul{q} , z_3) ,
\end{align}
which are calculated in the conventions of
\cite{Kovchegov:2012mbw}. Here and below LO stands for leading order,
NLO stands for next-to-leading order, etc. The virtual correction
$\mathcal{V}$ is obtained by unitarity, requiring that the sum of
$\ord{\alpha_s}$ corrections to the (anti)quark wave function not
modify the normalization:
\begin{align} 
  \label{e:virt}
  \mathcal{V} = \mathcal{V}^* = - \half
  \sum_{\mathrm{colors}}\sum_{\sigma'' \eta} \int \frac{d z_2}{4\pi
    z_2} \frac{d^2 k_2}{(2\pi)^2} | \psi_{LO}^{unp} (\ul{k}_2 , z_2)
  |^2 .
\end{align}
Using this to simplify \eqref{e:realvirt1}, we obtain
\begin{align}
  \mathcal{A}_{1 0} = - \int \!\!\frac{d z_3}{4 \pi z_3} d^2 &x_3
  \sum_{\mathrm{colors}} \sum_{\lambda \sigma'} \lambda \,
  \tilde\psi_{LO}^{pol} (\ul{x}_{31} , z_3) \times \frac{1}{z_3 s}
  \llangle U_{\ul 3}^{pol} \, V_1^\dagger \rrangle (z_3)
  \notag \\ &\times
  \left[ 2 \mathcal{V} \, \tilde\psi_{LO}^{unp} (\ul{x}_{31} , z_3) +
    \tilde\psi_{NLO}^{unp} (\ul{x}_{31} , z_3) \right]^* ,
\end{align}
and the real-virtual cancellations are contained within the sum in
brackets.

The first term is straightforward to calculate using \eqref{e:virt},
\begin{align} 
  \label{e:realvirt2}
  2 \mathcal{V} \, \tilde\psi_{LO}^{unp} (\ul{x}_{31} , z_3) &=
  \frac{i g^3 N_c}{(2\pi)^3} t^b \delta_{\sigma \sigma'} \:
  \frac{\ul{\epsilon}_\lambda^* \cdot \ul{x}_{32}}{x_{32}^2} \,
  \int\limits_{z_3}^{z_1} \frac{d z_2}{z_2} \int\limits_{1/z_2
    s}^{x_{10}^2 \tfrac{z_1}{z_2}} \frac{d x_{21}^2}{x_{21}^2} ,
\end{align}
where the DLA part comes from the regime $z_1 \gg z_2 \gg z_3$ and
$x_{10}^2 z_1 \gg x_{21}^2 z_2$, with no apparent further constraint
on $x_{32}$.  Note that the sign of \eqref{e:realvirt2} is important,
and one must correctly incorporate the sign for antiquark vertices
(see, e.g., Rule 3 following Eq. (3.28) of \cite{Brodsky:1997de}).  We
have also written the color factor $C_F \approx N_c / 2$ in the
large-$N_c$ limit.  Meanwhile, for the second term, we have in
momentum space
\begin{align}
  \psi_{NLO}^{unp} (\ul{q}, z_3) = - g^3 N_c t^b \:
  \frac{\ul{\epsilon}_\lambda^*\cdot\ul{q}}{q_T^2}
  \int\limits_{z_3}^{z_1} \frac{d z_2}{(z_2)^2} \int \frac{d^2
    k}{(2\pi)^3} \, \frac{\ul{k} \cdot (\ul{k} - \ul{q})}{k_T^2}
  \left( \frac{q_T^2}{2 z_3} + \frac{(k-q)_T^2}{2 z_2} \right)^{-1}
\end{align}
which is only DLA if $(k-q)_T^2 \gg \tfrac{z_2}{z_3} q_T^2 \gg q_T^2$.
In coordinate space, this corresponds to $x_{21}^2 \ll
\tfrac{z_3}{z_2} x_{32}^2 \ll x_{32}^2 \approx x_{31}^2$, giving the
DLA part as
\begin{align} 
  \label{e:realvirt3}
  \tilde\psi_{NLO}^{unp} (\ul{x}_{31}, z_3) = - \frac{ i g^3
    N_c}{(2\pi)^3} t^b \: \frac{\ul{\epsilon}_\lambda^*
    \cdot\ul{x}_{31}}{x_{31}^2} \int\limits_{z_3}^{z_1} \frac{d
    z_2}{z_2} \int\limits_{1/z_2 s}^{x_{10}^2 \tfrac{z_1}{z_2}}
  \frac{d x_{21}^2}{x_{21}^2} \, \theta(z_3 x_{31}^2 - z_2 x_{21}^2) .
\end{align}

We see that in the regime $z_3 x_{31}^2 \gg z_2 x_{21}^2$ (and for
$z_2 \gg z_3$), all three diagrams of $\mathcal{A}_{10}$ are DLA and
cancel so that $\mathcal{A}_{10} \approx 0$ with DLA accuracy.  This
means that the second step of evolution which produces the gluon
$\ul{x}_3$ is actually {\it{not}} DLA in the whole phase space; the
only DLA phase space which survives these cancellations is $z_3
x_{31}^2 \ll z_2 x_{21}^2$ (again, for $z_2 \gg z_3$).  Therefore,
when we write the BFKL-type virtual corrections as in the first line
of Fig.~\ref{f:real_virt_cancel} (or in the right two diagrams of the
second line of \fig{f:evol_step_latex}), we see that the subsequent
DLA evolution of the dipole 10 implicitly has the condition $z_3
x_{31}^2 \ll z_2 x_{21}^2$ imposed on it, so that the dipole amplitude
is not $G_{10} (z_2)$, but rather the {\it{neighbor dipole amplitude}}
$\Gamma_{10 , 21} (z_2)$.  This is the reason why the virtual
corrections (last terms of \eqref{e:largeNevol1} and
\eqref{e:largeNevol2}) enter with the neighbor dipole constraint on
their evolution.

\subsubsection{Virtual Corrections and Unitarity}

\begin{figure}[hbt]
 \centering
 \includegraphics[width=\textwidth]{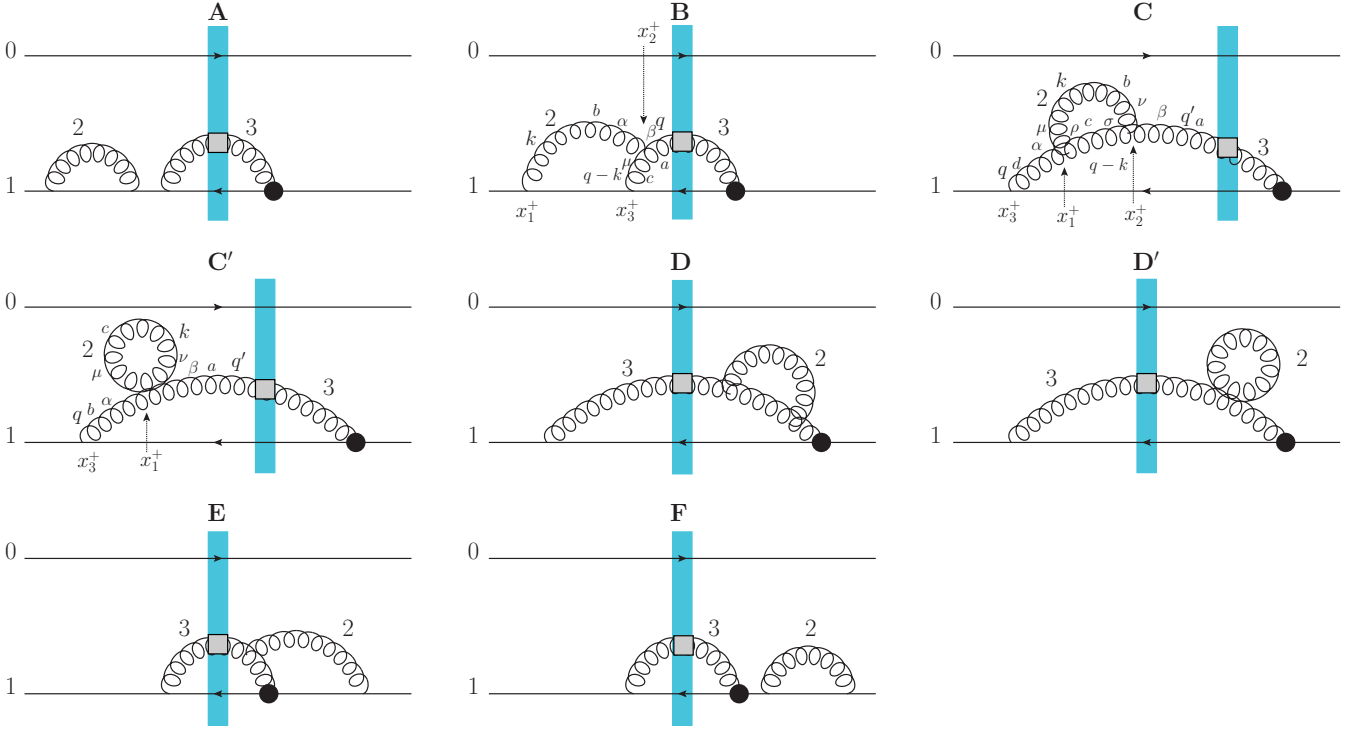}
 \caption{DLA diagrams in which a BFKL-like virtual correction follows
   the real emission of a soft polarized gluon.  We only consider
   diagrams which contribute to the further evolution of the $(01)$
   and $(03)$ dipoles.}
\label{f:real_virt_2}
\end{figure}

\begin{figure}[hbt]
 \centering
 \includegraphics[width=0.67\textwidth]{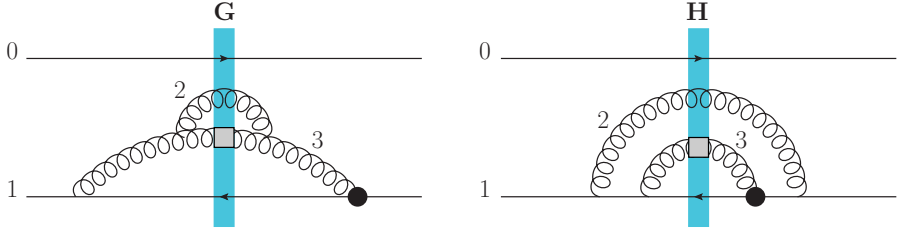}
 \caption{DLA diagrams in which a BFKL-like real correction follows
   the real emission of a soft polarized gluon.  We only consider
   diagrams which contribute to the further evolution of the $(01)$
   and $(03)$ dipoles.}
\label{f:real_virt_3}
\end{figure}

For completeness, let us study the case of opposite ordering, $z_3 \gg
z_2$.  Consider the two steps of DLA evolution in the opposite order:
first the emission of a soft polarized gluon 3, followed by a
BFKL-type correction (gluon 2) in the 01 / 03 dipoles included in all
possible ways.  In this case there are many possible virtual
corrections to consider (Fig.~\ref{f:real_virt_2}, diagrams $A$ --
$F$) and two real corrections (Fig.~\ref{f:real_virt_3}, diagrams $G$
-- $H$).  For ease of comparison with Fig.~\ref{f:real_virt_WF}, we
keep the polarized soft gluon to be at position $\ul{x}_3$ with
momentum $q$ and the BFKL-like unpolarized gluon to be at position
$\ul{x}_2$ with momentum $k$.  This ordering of the two evolution
steps then corresponds to $\tfrac{q^+}{k^+} = \tfrac{z_3}{z_2} \gg 1$.

There are two separate DLA regimes for the graphs in
Figs.~\ref{f:real_virt_2} and \ref{f:real_virt_3}, which are easily
understood in the language of LFPT.  The light-front energy (minus
momentum) $E_2 \, (E_3)$ of gluon $2 \, (3)$ is directly related to
the lifetimes of the gluon fluctuation; in coordinate space, these
energies are: $E_2 = \tfrac{1}{z_2 x_{21}^2}$ for diagrams $A$, $B$,
$E$, $F$, and $H$; $E_2 = \tfrac{1}{z_2 x_{23}^2}$ for diagrams $C$,
$C'$, $D$, $D'$, and $G$; and $E_3 = \tfrac{1}{z_3 x_{31}^2}$ for all
diagrams. (For brevity, in light-front energies we dropped the overall
factor of $1/p^+$ with $p^+$ the probe's momentum.)  As a rule of
thumb, the two steps of evolution shown here are DLA when there is a
well-separated hierarchy of lifetimes, such that the light-front
energy of each gluon $2$ and $3$ dominates exactly two of the
intermediate states, $E_2 \gg E_3$ or $E_3 \gg E_2$. (The application
of this rule gets more nuanced for diagrams with virtual corrections.)

For $E_2 \gg E_3$, all the diagrams in Figs.~\ref{f:real_virt_2} and
\ref{f:real_virt_3} can be DLA except for the real diagram $H$.  We
would like to compute the sum of the virtual diagrams $A$ -- $F$ in
order to assess the cancellations which may occur between them.  To do
this, we will calculate the relevant parts of diagrams $B$, $C$, and
$C'$; the analogous calculation of diagrams $D$, $D'$ and $E$ is
almost equivalent, the small difference being due to the position of
the polarized vertex to the right of the shock wave. It is convenient
to do these calculations in Feynman perturbation theory, rather than
in LFPT directly, treating the quark propagators as Wilson lines.  To
impose the corresponding time ordering, we need to Fourier transform
each gluon propagator from $k^-$ momentum space to $x^+$ coordinate
space, and then integrate over all light-cone ``times'' of the
vertices $x_i^+$ with the ordering prescribed by the diagram.

We can apply this scheme to just the parts of diagrams $B$, $C$, $C'$
which involve the radiation of gluons to the left of the shock wave;
everything else is common to the three diagrams.  Doing this, we
obtain
\begin{subequations}\label{BCC1}
 \begin{align}
   B = & \int\limits_{-\infty}^0 d x_1^+ \int\limits_{x_1^+}^0 d x_3^+
   \int\limits_{x_3^+}^0 d x_2^+ \, e^{\epsilon \, (x_1^+ + x_2^+ +
     x_3^+)} \int\limits_{-\infty}^{\infty} \frac{d k^-}{2 \pi} \,
   \frac{d q^-}{2 \pi} \, \frac{d (q-k)^-}{2 \pi} \,
   \\ &\times \, \notag
   e^{- i k^- (x_2^+ - x_1^+) - i (q-k)^- (x_2^+ - x_3^+) + i q^-
     x_2^+} \, \times \, \hat{B} ,
 \\ 
 C = & \int\limits_{-\infty}^0 d x_3^+ \int\limits_{x_3^+}^0 d x_1^+
 \int\limits_{x_1^+}^0 d x_2^+ \, e^{\epsilon \, (x_1^+ + x_2^+ +
   x_3^+)} \int\limits_{-\infty}^{\infty} \frac{d k^-}{2 \pi} \,
 \frac{d q^-}{2 \pi} \, \frac{d (q-k)^-}{2 \pi} \, \frac{d q'^-}{2
   \pi}
   \\ & \times \, \notag
   e^{- i k^- (x_2^+ - x_1^+) - i (q-k)^- (x_2^+ - x_1^+) + i q'^-
     x_2^+ - i q^- (x_1^+ - x_3^+)} \times \hat{C} ,
 \\ 
 C' = & \int\limits_{-\infty}^0 d x_3^+ \int\limits_{x_3^+}^0 d x_1^+
 \, e^{\epsilon \,( x_1^+ + x_3^+)} \int\limits_{-\infty}^{\infty}
 \frac{d k^-}{2 \pi} \, \frac{d q^-}{2 \pi} \, \frac{d q'^-}{2 \pi} \,
 e^{i q'^- x_1^+ - i q^- (x_1^+ - x_3^+)} \times \hat{C}' .
\end{align}
\end{subequations}
In \eqref{BCC1}, we use the regulator $e^{\epsilon x^+}$ for positive
infinitesimal $\epsilon$ to ensure convergence at $x^+ \rightarrow
-\infty$ (see e.g.~\cite{Chirilli:2015tea}).  The shock wave
(interaction with the target) is taken to occur at $x^+ = 0^+$.  The
momentum-space expressions, calculated in Feynman perturbation theory,
are
\begin{subequations} \label{BCC2}
 \begin{align}
   \hat{B} &= (-i g)^2 \, g \, f^{abc} t^b t^c \left( \frac{-i}{k^2 +
       i \delta} \right) \left( \frac{-i}{(q-k)^2 + i \delta} \right)
   \left( \frac{i \, \epsilon_\lambda^{* \, \beta} (q)}{q^{2} + i
       \delta} \right) \left[ g^{-\alpha} - \frac{k^\alpha + g^{+
         \alpha} k^-}{k^+} \right]
   \notag \\ & \times \, 
   \left[ g^{-\mu} - \frac{(q-k)^\mu + g^{+ \mu} (q-k)^-}{q^+ - k^+}
   \right] \Big[ (2 k - q)_\beta \, g_{\mu\alpha} - (q+k)_\mu \,
   g_{\alpha \beta} + (2 q -k)_\alpha \, g_{\mu\beta} \Big] ,
 \\ \notag \\
 \hat{C} &= \frac{-i g^3}{2} \, f^{abc} \, f^{cbd} \, t^d \left(
   \frac{-i}{k^2 + i \delta} \right) \left( \frac{-i}{q^2 + i \delta}
 \right) \left( \frac{-i}{(q-k)^2 + i \delta} \right) \left( \frac{i
     \, \epsilon_\lambda^{* \, \beta} (q')}{q'^{\, 2} + i \delta}
 \right) 
   \notag \\ & \times \,
   \left[ g^{-\alpha} - \frac{q^\alpha + g^{+ \alpha} q^-}{q^+}
   \right] \left[ g^{\mu\nu} - \frac{k^\mu g^{+ \nu} + k^\nu g^{+ \mu}
     }{k^+} \right] \left[ g^{\rho\sigma} - \frac{(q-k)^\rho g^{+
         \sigma} + (q-k)^\sigma g^{+ \rho}}{q^+ - k^+} \right]
   \notag \\ & \times \,
   \Big[ (2 k - q)_\alpha \, g_{\mu\rho} - (q+k)_\rho \, g_{\mu\alpha}
   + (2 q -k)_\mu \, g_{\alpha\rho} \Big]
   \notag \\ & \times \,  
   \Big[ (2 k - q)_\beta \, g_{\sigma\nu} - (q'+k)_\sigma \, g_{\nu
     \beta} + (q + q' -k)_\nu \, g_{\sigma\beta} \Big] ,
 \\ \notag \\
 \hat{C}' &= (-g^3) \, \frac{N_c}{2} \, t^a \left( \frac{-i}{q^2 + i
     \delta} \right) \left( \frac{-i}{k^2 + i \delta} \right) \left(
   \frac{i \, \epsilon_\lambda^{* \, \beta} (q')}{q'^{\, 2} + i
     \delta} \right) \left[ g^{-\alpha} - \frac{q^\alpha + g^{+
       \alpha} q^-}{q^+} \right] \,
   \notag \\ & \times \, 
   \left[ g^{\mu\nu} - \frac{g^{+ \nu} k^\mu + g^{+ \mu} k^\nu }{k^+}
   \right] \, \left[2 \, g_{\mu\nu} \, g_{\alpha\beta} - g_{\mu\beta}
     \, g_{\nu\alpha} - g_{\mu\alpha} \, g_{\nu\beta} \right].
 \end{align}
\end{subequations}
In \eqref{BCC2}, we use $i\delta$ for the regulator of the Feynman
propagator, and we have split (the numerator of) the propagator of
gluon 3 through the shock wave into a polarization sum, keeping only
the gluon polarization $\epsilon_\lambda^*$.  Note that diagrams $C$
and $C'$ come with an explicit minus sign due to the antiquark/gluon
vertex (see \eqref{e:realvirt2} and the discussion thereafter) and
that only half of their color factor ``belongs'' to the evolution of
dipole 03 under consideration, the other half being the evolution in
the dipole 31 which we do not consider here.

Keeping only the leading-energy, DLA part of the expressions, we
obtain\footnote{Indeed diagram $C'$ contains a UV divergence, which
  has to be canceled by a counter-term. This contribution is not DLA
  and is not shown in Eqs.~\eqref{BCC3}.}
\begin{subequations}\label{BCC3}
 \begin{align}
   B = & \, g^3 \, \frac{N_c}{2} \, t^a \, \frac{1}{2 \, k^+ \, q^+ \,
     k_\perp^2 \, q_\perp^2} \, \left[ \, \ul{\epsilon}_\lambda^*
     \cdot (- 4 \ul{q} + 2 \ul{k}) \right],
   \\
   C = & \, g^3 \, \frac{N_c}{2} \, t^a \, \frac{1}{2 \, k^+ \, q^+ \,
     k_\perp^2 \, q_\perp^2} \, \left[ \ul{\epsilon}_\lambda^* \cdot
     (4 \, \ul{q} - 2 \, \ul{k}) \right],
   \\
   C' = & \, 0 .
 \end{align}
\end{subequations}
Thus we see that, with DLA accuracy, $B + C + C' = 0$.  By an
analogous calculation, one also finds that $D + D' + E = 0$.  The
result is that, for $E_2 \gg E_3$, only the virtual diagrams $A \, ,
\, F$ from Fig.~\ref{f:real_virt_2} and the real diagram $G$ from
Fig.~\ref{f:real_virt_3} contribute to the DLA evolution of dipole 01
followed by the LLA-type evolution of dipole 03. Note that the latter
LLA-type step comes with the $E_2 \gg E_3$ condition, normally not
associated with the LLA evolution.

On the other hand, in the $E_3 \gg E_2$ regime, only the virtual
diagrams $A \, , \, F$ from Fig.~\ref{f:real_virt_2} and the real
diagram $H$ from Fig.~\ref{f:real_virt_3} are DLA.  In this kinematic
regime, we have $x_{21}^2 \gg \tfrac{z_3}{z_2} x_{31}^2 \gg x_{31}^2$
so that the dipole $31$ is very small: gluon 3 is very close to the
parent antiquark 1.  Diagram $H$ becomes indistinguishable from
diagram $G$, since gluon 2 is essentially emitted from coordinate
$x_3$ in both cases.  Diagrams $A$, $F$ and $H$ then contribute to
LLA-type evolution in the dipole $01 \approx 03$, now with the $E_3
\gg E_2$ condition. Combining this with the contributions of diagrams
$A$, $F$ and $G$ in the $E_2 \gg E_3$ regime we obtain LLA evolution
in the dipole 03 without any ordering of the light-cone energies, as
is normal for the LLA evolution. Such contribution is included in
Eqs.~\eqref{e:largeNevol}.

The result of this analysis is that, in either DLA limit $E_2 \gg E_3$
or $E_3 \gg E_2$, one is left only with the virtual corrections $A \,
, \, F$ and the (equivalent) real correction $G / H$.  These BFKL-like
real and virtual corrections are exactly the ones included in our
evolution equation \eqref{e:opevol1} and Fig.~\ref{f:evol_step_latex}.
And, moreover, in the absence of {\it{any}} interactions with the shock
wave, $V , V^{pol} = 1$, these real and virtual corrections cancel
exactly (see \eqref{e:virt}), as demanded by unitarity.  Therefore we
conclude that our treatment of BFKL-like virtual corrections exhausts
the unitarity sum (with DLA accuracy), completing the cross-check of
the way we have implemented these {\it{unpolarized}} radiative
corrections to the polarized dipole amplitude.


\subsection{Cross-Check: DGLAP Anomalous Dimensions}
\label{sec:DGLAP}

Another important cross-check of our evolution equations is to verify
that they reproduce the correct DGLAP anomalous dimensions at NLO
accuracy.  This is especially important in reconciling our
disagreement with BER, since in Eq.~(4.25) of their work
\cite{Bartels:1996wc}, BER show that they reproduce the complete LO
and NLO DGLAP polarized anomalous dimensions. (In addition, it was
recently shown that the result of BER's formalism, expanded to higher
orders in \cite{Blumlein:1996hb}, correctly reproduces the NNLO
polarized anomalous dimensions \cite{Moch:2014sna}.)  For BER,
obtaining anomalous dimensions is a straightforward application of
their infrared evolution equations which re-sum the mixed logarithms
$\alpha_s^i (\ln \tfrac{1}{x})^{2i - j} (\ln \tfrac{Q^2}{\mu^2})^j$
for $0 \leq j \leq i$, such that their final answer contains all-order
small-$x$ anomalous dimensions for DGLAP evolution. One simply needs
to expand this anomalous dimension to order $\as^2$ to obtain the
small-$x$ contribution to the NLO anomalous dimension.

In our case, the correspondence is less clear, chiefly because, unlike
BER, we do not have an exact analytic solution for our evolution
equations and our evolution only resums powers of $\as \ln^2
\tfrac{1}{x}$. In addition, our equations do not close in general [see
\eqref{e:opevol1}], and, hence, cannot be used to easily extract the
anomalous dimension of any of the involved operators. However, our
large-$N_c$ equations (in the flavor singlet case) close. Moreover,
they can be written as a single closed equation for the expectation
value of only one operator. Noting that the integrands are the same in
\eqref{e:linear}, we formulate the evolution equations in terms of the
linear combination
 \begin{align}
   H(x_T^2 , y_T^2 , z) &\equiv \Gamma(x_T^2 , y_T^2 , z) + 3 G(y_T^2
   , z) ,
 \end{align}
 giving
 \begin{align} 
   \label{e:Hevol}
   H(x_{10}^2 , x_{21}^2 , z) &= G^{(0)} (x_{10}^2 , z) + 3 G^{(0)}
   (x_{21}^2 , z) + \frac{\alpha_s N_c}{2\pi}
   \int\limits_{\tfrac{1}{x_{10}^2 s}}^{z} \frac{d z'}{z'} \!\!
   \int\limits_{\tfrac{1}{z' s}}^{\min\left[x_{10}^2 \, , \, x_{21}^2
       \tfrac{z}{z'}\right]} 
   \frac{d x_{32}^2}{x_{32}^2} \, H(x_{10}^2 , x_{32}^2 , z')
   \notag \\ &+
   3 \, \frac{\alpha_s N_c}{2\pi} \int\limits_{\tfrac{1}{x_{21}^2
       s}}^{z} \frac{d z'}{z'} \int\limits_{\tfrac{1}{z'
       s}}^{x_{21}^2} \frac{d x_{32}^2}{x_{32}^2}\, H (x_{21}^2 ,
   x_{32}^2 , z') .
\end{align}
The resulting \eq{e:Hevol} contains only gluon bremsstrahlung, so we
only have access to the glue-glue sector of the splitting kernel in
the large-$N_c$ approximation. (Our flavor-singlet helicity evolution
equations also close in the large-$N_c \& N_f$ limit
\cite{Kovchegov:2015pbl}; however, the resulting closed equations
depend on the expectation values of several operators. We leave it for
the future work to elucidate the possibility of extracting small-$x$
polarized NLO DGLAP anomalous dimensions in the quark-quark,
quark-gluon and gluon-quark sectors from those equations.)

DGLAP evolution expresses a PDF $f_i (x, Q^2)$ at one (UV) scale $Q^2$
and momentum fraction $x$ in terms of a convolution of PDF's at lower
(IR) scales $\mu^2 < k_T^2 < Q^2$ and higher momentum fractions $x'
\geq x$:
\begin{align} 
  \label{e:DGLAP1}
  f_i (x, Q^2) &= f_i (x, \mu^2) + \int\limits_{x}^{1} \frac{d x'}{x'}
  \int\limits_{\mu^2}^{Q^2} \frac{d k_T^2}{k_T^2} \, P_{i / j}
  (\tfrac{x}{x'}) \, f_j (x' , k_T^2).
\end{align}
The splitting functions are expanded in a perturbation series in $\as$, 
\begin{align}
  \label{eq:split_exp}
  P_{i / j} (z) = P_{i / j}^{LO} (z) + P_{i / j}^{NLO} (z) + \ldots ,
\end{align}
where the $LO$ term is ${\cal O} (\as)$, the NLO term is ${\cal O}
(\as^2)$, etc.

Our integral evolution equations, however, express the evolution ``in
the opposite direction'' to standard DGLAP evolution.  They express a
polarized dipole distribution at one (IR) scale $\mu^2$ and momentum
fraction $x$ in terms of a convolution of dipole distributions at
{\it{higher (UV)}} scales $\mu^2 < \tfrac{1}{x_{21}^2} < Q^2$ and
{\it{lower}} momentum fractions $x' \leq x$.  For example,
\begin{align}
  G(\tfrac{1}{\mu^2} , x) &= G ^{(0)} (\tfrac{1}{\mu^2} , x) +
  \frac{\alpha_s N_c}{2\pi} \int\limits_{\tfrac{\mu^2}{s}}^{x}
  \frac{d x'}{x'} \int\limits_{\left(\tfrac{x}{x'}\right)
    \tfrac{1}{Q^2}}^{\tfrac{1}{\mu^2}} \frac{d x_{21}^2}{x_{21}^2}
  \left[ \Gamma(\tfrac{1}{\mu^2} , x_{21}^2 , x') + 3 G(x_{21}^2 , x')
  \right] ,
\end{align}
where we take $x = \tfrac{Q^2}{s}$ as in deep inelastic scattering at
small $x$.

Clearly it would be difficult to recast our evolution equations into a
form which can be easily compared to DGLAP. However, the
{\it{kernels}} or {\it{splitting functions}} of the two equations
should be comparable with one another, since they are built at the
fundamental level from the same ingredients:~the light-front splitting
wave functions of quarks and gluons.  Our strategy, then, will be to
iterate our equation \eqref{e:Hevol} to the desired order, since it is
a closed equation for a single function $H$, translate the result into
DGLAP kinematics, and then extract the splitting function from it.

Our DLA evolution equations generate two logarithms of energy after
each step of evolution is completely integrated; when these double
logarithms of energy are translated from helicity evolution in Regge
kinematics to DGLAP evolution in Bjorken kinematics, some of them will
correspond to $\alpha_s \ln\tfrac{Q^2}{\mu^2}\ln\tfrac{1}{x}$ and
others will correspond to $\alpha_s \ln^2\tfrac{1}{x}$.  The former
category of terms is leading-logarithmic in $Q^2$ and thus contributes
to the LO DGLAP anomalous dimension.  The latter category of terms is
subleading in $Q^2$ and thus suppressed in the DGLAP hierarchy
$\ln\tfrac{Q^2}{\mu^2} \gg \ln\tfrac{1}{x}$.  The NLO anomalous
dimension, therefore, comes from terms of order $\alpha_s^2
\ln\tfrac{Q^2}{\mu^2} \ln^3\tfrac{1}{x}$ and requires two iterations
of our evolution equation to compute.  Terms which contain no
logarithms of $Q^2$, that is $(\alpha_s \ln^2\tfrac{1}{x})^n$,
contribute to our evolution equation but not DGLAP evolution.  Note
that one logarithm of $\tfrac{1}{x}$ and one logarithm of $Q^2$ are
contained in the explicit integral in \eqref{e:DGLAP1}, so that the
NLO terms of interest in the splitting function $P$ are of order
$\alpha_s^2 \ln^2\tfrac{1}{x}$.

There is one further complication which is specific to our evolution
equations: due to the ``neighbor dipole'' functions, the amplitude $H$
which evolves in \eqref{e:Hevol} depends on two scales ($x_{01}^2$ and
$x_{21}^2$) rather than one like the PDF's and the DGLAP kernel. The
reason behind this is that the neighbor dipole ``remembers'' one of
the previous evolution steps. This makes it impossible to identify two
steps of our evolution simply with two gluon emissions: the neighbor
dipole takes into account at least one previous gluon emission. The
problem here is in separating the NLO contribution coming from the
two-gluon emission generated by two steps of helicity evolution (the
result we want) from the admixture of the NLO contribution coming from
the earlier gluon emission.

\begin{figure}[ht]
 \begin{center} 
  \includegraphics[width=0.6\textwidth]{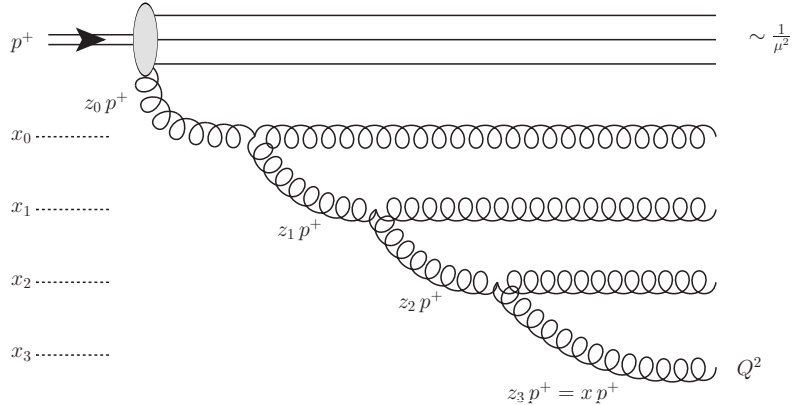} 
  \caption{Three gluon emissions, representing one step of the LO
    DGLAP-type evolution, followed by two steps of our DLA evolution,
    as suggested in \eq{eq:plan}.}
  \label{f:glue}
 \end{center}
\end{figure}

The neighbor dipole is not directly observable; it only influences the
evolution of the observable quantity $G(\tfrac{1}{\mu^2} , z_0)$.  We
could eliminate the above mentioned ambiguity by directly performing
the first step of evolution in which the neighbor dipoles are
generated,
\begin{align} 
  \label{e:Gstep}
  G(\tfrac{1}{\mu^2} , z_0) = \frac{\alpha_s N_c}{2\pi}
  \int\limits_{\tfrac{\mu^2}{s}}^{z_0} \frac{d z_1}{z_1}
  \int\limits_{\tfrac{1}{z_1 s}}^{\tfrac{1}{\mu^2}} \frac{d
    x_{01}^2}{x_{01}^2} \, H(\tfrac{1}{\mu^2} , x_{01}^2 , z_1),
\end{align}
where we have neglected the initial conditions, since they do not
generate double logarithms.  Then, performing two steps of $H$
evolution by iterating \eqref{e:Hevol}, with all of the resulting
emissions shown in Fig.~\ref{f:glue}, we obtain
\begin{align}
  \label{steps23}
  & G(\tfrac{1}{\mu^2} , z_0) = \left( \frac{\alpha_s N_c}{2\pi}
  \right)^3 \, \int\limits_{\tfrac{\mu^2}{s}}^{z_0} \frac{d z_1}{z_1}
  \int\limits_{\tfrac{\mu^2}{s}}^{z_1} \frac{d z_2}{z_2}
  \int\limits_{\tfrac{\mu^2}{s}}^{z_2} \frac{d z_3}{z_3}
  \int\limits_{\tfrac{1}{z_1 s}}^{\tfrac{1}{\mu^2}} \frac{d
    x_{01}^2}{x_{01}^2}
	\notag \\ & \times \, 
	\left\{ \ \int\limits_{\tfrac{1}{z_2 s}}^{\min \left[
              \tfrac{1}{\mu^2}, \tfrac{x_{01}^2 z_1}{z_2}\right]}
          \frac{d x_{21}^2}{x_{21}^2} \left[
            \int\limits_{\tfrac{1}{z_3 s}}^{\min \left[
                \tfrac{1}{\mu^2}, \tfrac{x_{21}^2 \, z_2}{z_3}\right]}
            \frac{d x_{32}^2}{x_{32}^2} \, H(\tfrac{1}{\mu^2} ,
            x_{32}^2 , z_3)
	+
	3 \, \theta ( x_{21}^2 - \tfrac{1}{z_3 s} ) \,
        \int\limits_{\tfrac{1}{z_3 s}}^{x_{21}^2} \frac{d
          x_{32}^2}{x_{32}^2} \, H(x_{21}^2 , x_{32}^2 , z_3) \right]
    \right.
	\notag \\ & \left. + \, 
          3 \, \theta ( x_{01}^2 - \tfrac{1}{z_2 s} ) \,
          \int\limits_{\tfrac{1}{z_2 s}}^{x_{01}^2} \frac{d
            x_{21}^2}{x_{21}^2} \left[ \int\limits_{\tfrac{1}{z_3
                s}}^{\min \left[ x_{01}^2, \tfrac{x_{21}^2 \,
                  z_2}{z_3}\right]} \frac{d x_{32}^2}{x_{32}^2} \,
            H(x_{01}^2 , x_{32}^2 , z_3)
	+
	3 \, \theta ( x_{21}^2 - \tfrac{1}{z_3 s} ) \,
        \int\limits_{\tfrac{1}{z_3 s}}^{x_{21}^2} \frac{d
          x_{32}^2}{x_{32}^2} \, H(x_{21}^2 , x_{32}^2 , z_3) \right]
    \right\}.
\end{align}
Since we are interested only in two steps of DLA $H$ evolution, we can
replace $H \to 1$ to neglect further evolution in \eq{steps23} and
recast it into the form
\begin{align}
  G (\tfrac{1}{\mu^2} , z_0) = \int\limits_{\tfrac{\mu^2}{s}}^{z_0}
  \frac{d z_3}{z_3} \, \mathcal{K}_{[DLA^3]} (\tfrac{z_3}{z_0},
    \tfrac{z_3 s}{\mu^2} ).
\end{align}
We see that our evolution equation, like DGLAP evolution, can be
expressed in the form of a convolution over a splitting kernel
$\mathcal{K} (\tfrac{z_3}{z_0}, \tfrac{s}{\mu^2})$.  Although our
equations themselves are not comparable to DGLAP, this splitting
kernel is.  To make the comparison more explicit, we can analyze the
kernel for fixed $z_3 = x = \tfrac{Q^2}{s}$, as appropriate for deep
inelastic scattering at small $x$. We then have
\begin{align}
  \mathcal{K}_{[DLA^3]} (\tfrac{x}{z_0}, \tfrac{Q^2}{\mu^2} )
  &= \left( \frac{\alpha_s N_c}{2\pi} \right)^3 \,
  \int\limits_{x}^{z_0} \frac{d z_1}{z_1} \int\limits_{x}^{z_1}
  \frac{d z_2}{z_2} \int\limits_{\tfrac{x}{z_1 \,
      Q^2}}^{\tfrac{1}{\mu^2}} \frac{d x_{01}^2}{x_{01}^2}
	\notag \\ & \times \, 
	\left\{ \ \int\limits_{\tfrac{x}{z_2 \, Q^2}}^{\min \left[
              \tfrac{1}{\mu^2}, \tfrac{x_{01}^2 z_1}{z_2}\right]}
          \frac{d x_{21}^2}{x_{21}^2} \left[
            \int\limits_{\tfrac{1}{Q^2}}^{\min \left[
                \tfrac{1}{\mu^2}, \tfrac{x_{21}^2 \, z_2}{x}\right]}
            \frac{d x_{32}^2}{x_{32}^2}
	+
	3 \, \theta ( x_{21}^2 - \tfrac{1}{Q^2} ) \,
  \int\limits_{\tfrac{1}{Q^2}}^{x_{21}^2} \frac{d x_{32}^2}{x_{32}^2} 
	\right] \right. 
	\notag \\ & \left. + \, 
	3 \, \theta ( x_{01}^2 - \tfrac{x}{z_2 \, Q^2} ) \,  
	\int\limits_{\tfrac{x}{z_2 \, Q^2}}^{x_{01}^2} \frac{d x_{21}^2}{x_{21}^2} 
	\left[ \int\limits_{\tfrac{1}{Q^2}}^{\min \left[ x_{01}^2, 
	\tfrac{x_{21}^2 \, z_2}{x}\right]} \frac{d x_{32}^2}{x_{32}^2} 
	+
	3 \, \theta ( x_{21}^2 - \tfrac{1}{Q^2} ) \,
  \int\limits_{\tfrac{1}{Q^2}}^{x_{21}^2} \frac{d x_{32}^2}{x_{32}^2} 
	\right] \right\}.
\end{align}

There are, unfortunately, a couple of problems with directly
extracting the NLO anomalous dimensions from this kernel.  Foremost,
the first step \eqref{e:Gstep} is not a ``diagonal'' evolution like
DGLAP: it connects two different functions $G$ and $H$.  Moreover, as
we have suggested by labeling the kernel $DLA^3$, the integrals in
this first step could potentially generate contributions to the NLO
anomalous dimension, which would contaminate the ``true'' NLO kernel
generated during the two steps of DLA evolution of $H$.  We can
resolve this problem by restricting this first step of evolution to
the LO DGLAP phase space; that is, we can impose the strict transverse
ordering $\tfrac{1}{Q^2} < x_{01}^2 < \tfrac{1}{\mu^2}$ in
\eq{e:Gstep}.  Thus, we can compute the modified kernel
$\mathcal{K}_{[LO \times DLA^2]}$ for {\it{three}} steps of evolution:
\begin{align}
  \label{eq:plan}
  G \overset{LO}{\Longrightarrow} H \overset{DLA}{\Longrightarrow} H
  \overset{DLA}{\Longrightarrow} H.
\end{align}
This eliminates the scale ambiguity coming from the neighbor dipoles,
and it guarantees that the NLO anomalous dimension we compute arises
purely from the ``diagonal'' evolution of $H$ and can be connected to
the NLO DGLAP anomalous dimension. The three steps of evolution
\eqref{eq:plan} are illustrated diagrammatically in \fig{f:glue}.

For completeness, we rewrite the modified kernel explicitly,
\begin{align} 
  \label{e:Kmod}
  \mathcal{K}_{[LO \times DLA^2]} (\tfrac{x}{z_0},
    \tfrac{Q^2}{\mu^2} ) &= \left( \frac{\alpha_s N_c}{2\pi}
  \right)^3 \, \int\limits_{x}^{z_0} \frac{d z_1}{z_1}
  \int\limits_{x}^{z_1} \frac{d z_2}{z_2}
  \int\limits_{\tfrac{1}{Q^2}}^{\tfrac{1}{\mu^2}} \frac{d
    x_{01}^2}{x_{01}^2}
	\notag \\ & \times \, 
	\left\{ \
  \int\limits_{\tfrac{x}{z_2 \, Q^2}}^{\min \left[ \tfrac{1}{\mu^2},
  \tfrac{x_{01}^2 z_1}{z_2}\right]} \frac{d x_{21}^2}{x_{21}^2}
	\left[ \int\limits_{\tfrac{1}{Q^2}}^{\min \left[ \tfrac{1}{\mu^2},
	\tfrac{x_{21}^2 \, z_2}{x}\right]} \frac{d x_{32}^2}{x_{32}^2} 
	+
	3 \, \theta ( x_{21}^2 - \tfrac{1}{Q^2} ) \,
  \int\limits_{\tfrac{1}{Q^2}}^{x_{21}^2} \frac{d x_{32}^2}{x_{32}^2} 
	\right] \right. 
	\notag \\ & \left. + \, 
	3 \,
	\int\limits_{\tfrac{x}{z_2 \, Q^2}}^{x_{01}^2} \frac{d x_{21}^2}{x_{21}^2} 
	\left[ \int\limits_{\tfrac{1}{Q^2}}^{\min \left[ x_{01}^2, 
	\tfrac{x_{21}^2 \, z_2}{x}\right]} \frac{d x_{32}^2}{x_{32}^2} 
	+
	3 \, \theta ( x_{21}^2 - \tfrac{1}{Q^2} ) \,
        \int\limits_{\tfrac{1}{Q^2}}^{x_{21}^2} \frac{d
          x_{32}^2}{x_{32}^2} \right] \right\},
\end{align}
and we carry out the integrations employing DGLAP kinematics $\ln
\tfrac{Q^2}{\mu^2} \gg \ln \tfrac{1}{x}$ to neglect the theta-function
terms like $\theta ( \tfrac{\mu^2}{Q^2} - \tfrac{x}{z_1}
)$, $\theta ( \tfrac{\mu^2}{Q^2} - \tfrac{z_2}{z_1}
)$, etc. Such terms do not contribute to DGLAP evolution. After
$z$-integrations they give terms proportional $\theta (
  \tfrac{\mu^2}{Q^2} - x )$ which may be identified as a
higher-twist effect (see Sec.~\ref{sec:Conclusions} below for a
further discussion of these terms). We arrive at
\begin{align}
  \label{e:steps23b}
  \mathcal{K}_{[LO \times DLA^2]} (\tfrac{x}{z_0},
    \tfrac{Q^2}{\mu^2} ) = \left( \frac{\alpha_s N_c}{2\pi}
  \right)^3 \, \left[ \frac{4}{3} \, \ln^3 \frac{Q^2}{\mu^2} \, \ln^2
    \frac{z_0}{x} + \frac{2}{3} \, \ln^2 \frac{Q^2}{\mu^2} \, \ln^3
    \frac{z_0}{x} + \ldots \right],
\end{align}
where the ellipsis denote the terms with one or no logarithms of
$Q^2$, which are not important for DGLAP evolution at the order of
interest.

To claim that our kernel \eqref{e:steps23b} is comparable with the
DGLAP kernel, we need to explicitly make the connection with the LO
and NLO DGLAP splitting functions.  Only the diagonal $H
\overset{DLA}{\Longrightarrow} H \overset{DLA}{\Longrightarrow} H$ is
compatible with DGLAP, containing both $LO^2$ and $NLO$ contributions
depending on the logarithms.  That is, we claim that our kernel is
related to DGLAP by
\begin{align}
  \mathcal{K}_{[LO \times DLA^2]} (\tfrac{x}{z_0},
    \tfrac{Q^2}{\mu^2} ) &= \frac{\alpha_s N_c}{2\pi}
  \int\limits_x^{z_0} \frac{d z_1}{z_1}
  \int\limits_{\tfrac{1}{Q^2}}^{\tfrac{1}{\mu^2}} \frac{d
    x_{01}^2}{x_{01}^2} \left[ \mathcal{K}_{[LO^2]}^{DGLAP}
    (\tfrac{x}{z_1}, \, x_{01}^2 Q^2 ) +
    \mathcal{K}_{[NLO]}^{DGLAP} (\tfrac{x}{z_1}, \, x_{01}^2 Q^2
    ) + \ldots \right]
 \notag \\ &=
 \frac{\alpha_s N_c}{2\pi} \int\limits_x^{z_0} \frac{d z_1}{z_1}
 \int\limits_{\tfrac{1}{Q^2}}^{\tfrac{1}{\mu^2}} \frac{d
   x_{01}^2}{x_{01}^2} \left[ \int\limits_x^{z_1} \frac{d z_2}{z_2}
   \int\limits_{\tfrac{1}{Q^2}}^{x_{01}^2} \frac{d x_{32}^2}{x_{32}^2}
   \, \Delta P_{S, \, G G}^{LO} (\tfrac{x}{z_2} \ll 1)
   \int\limits_{x_{32}^2}^{x_{01}^2} \frac{d x_{21}^2}{x_{21}^2} \,
   \Delta P_{S, \, G G}^{LO} (\tfrac{z_2}{z_1} \ll 1) \right.
 \notag \\ & \hspace{2cm} +
 \left. \int\limits_{\tfrac{1}{Q^2}}^{x_{01}^2} \frac{d x_{31}^2}{x_{31}^2}
 \, \Delta P_{S, \, G G}^{NLO} (\tfrac{x}{z_1} \ll 1) \right]
\end{align}
where $\Delta P_{S, \, G G}^{LO}$ and $\Delta P_{S, \, G G}^{NLO}$ are
the LO and NLO polarized glue/glue splitting functions, respectively.

To check this assertion, we can directly compute the $LO^2$ part of
the splitting kernel, remembering that the LO glue-glue splitting
function for polarized DGLAP evolution is $\Delta P_{S, \, G G}^{LO}
(z\to 0) = 4 N_c (\as/2\pi)$ \cite{Altarelli:1977zs} (which can also
be derived from our \eq{e:Hevol}).  We obtain
\begin{align}
  \frac{\alpha_s N_c}{2\pi} \int\limits_x^{z_0} \frac{d z_1}{z_1}
  \int\limits_{\tfrac{1}{Q^2}}^{\tfrac{1}{\mu^2}} \frac{d
    x_{01}^2}{x_{01}^2} \mathcal{K}_{[LO^2]}^{DGLAP} (
    \tfrac{x}{z_1}, \, x_{01}^2 Q^2 ) = \left(\frac{\alpha_s
      N_c}{2\pi}\right)^3 \left[ \frac{4}{3} \ln^3 \frac{Q^2}{\mu^2}
    \ln^2 \frac{z_0}{x} \right].
\end{align}
We see that, indeed, this reproduces the $\ord{\ln^3 Q^2}$ term of
\eqref{e:steps23b}.  This is an important cross-check of our
calculation.  Subtracting off this $LO^3$ piece,
\begin{align}
  \mathcal{K}_{[LO \times DLA^2 - LO^3]} (\tfrac{x}{z_0},
    \tfrac{Q^2}{\mu^2} ) &\equiv \mathcal{K}_{[LO \times DLA^2]}
  (\tfrac{x}{z_0}, \tfrac{Q^2}{\mu^2} ) - \frac{\alpha_s
    N_c}{2\pi} \int\limits_x^{z_0} \frac{d z_1}{z_1}
  \int\limits_{\tfrac{1}{Q^2}}^{\tfrac{1}{\mu^2}} \frac{d
    x_{01}^2}{x_{01}^2} \: \mathcal{K}_{[LO^2]}^{DGLAP} (
    \tfrac{x}{z_1}, \, x_{01}^2 Q^2 )
 \notag \\ &=
 \frac{\alpha_s N_c}{2\pi} \int\limits_x^{z_0} \frac{d z_1}{z_1} 
 \int\limits_{\tfrac{1}{Q^2}}^{\tfrac{1}{\mu^2}} \frac{d x_{01}^2}{x_{01}^2}
 \int\limits_{\tfrac{1}{Q^2}}^{x_{01}^2} \frac{d x_{31}^2}{x_{31}^2}
 \, \Delta P_{S, \, G G}^{NLO} (\tfrac{x}{z_1} \ll 1) ,
\end{align}
we can extract the NLO splitting function from \eqref{e:steps23b} by
differentiating,
\begin{align}
  \Delta P_{S, \, G G}^{NLO} ( \tfrac{x}{z_0} \ll 1 ) =
  \frac{2\pi}{\alpha_s N_c} \frac{\partial}{\partial \ln
    \tfrac{z_0}{x}} \left( \frac{\partial} { \partial \ln
      \tfrac{Q^2}{\mu^2} } \right)^2 \, \mathcal{K}_{[LO \times DLA^2
    - LO^3]} (\tfrac{x}{z_0}, \tfrac{Q^2}{\mu^2} ) .
\end{align}
Doing so, we obtain
\begin{align}
  \label{Yglue}
  \Delta P_{S, \, GG}^{NLO} (z \to 0) \bigg|_{\mbox{pure} \,
    \mbox{glue}}= \left( \frac{\as}{2 \, \pi} \right)^2 4 \, N_c^2 \,
  \ln^2 z
\end{align}
in complete agreement with the literature \cite{Mertig:1995ny}.

The corresponding anomalous dimension can be found using
\begin{align}
  \label{eq:anom}
  \gamma (\omega) = \int\limits_0^1 \, dz \, z^{\omega -1} \, \Delta
  P(z).
\end{align}
We obtain the pure-glue flavor-singlet anomalous dimension
\begin{align}
  \gamma_{S, \, GG}^{NLO} (\omega) \bigg|_{\mbox{pure} \, \mbox{glue},
    \ \omega \to 0} = \left( \frac{\as}{2 \, \pi} \right)^2 \, \frac{8
    \, N_c^2}{\omega^3},
\end{align}
also in agreement with \cite{Mertig:1995ny}. 

We conclude that our helicity evolution generates the small-$x$
flavor-singlet polarized DGLAP glue-glue splitting function and
anomalous dimension, which are in complete agreement with the existing
LO \cite{Altarelli:1977zs} and NLO calculations \cite{Mertig:1995ny}.


\section{Flavor Non-Singlet Helicity Evolution}
\label{sec:FNS}

\subsection{Flavor Non-Singlet Initial conditions}

\label{FNSdef}

Let us now derive the evolution equations governing the small-$x$
behavior of the flavor non-singlet helicity distribution
\begin{align}
  \Delta q^{NS} (x, Q^2) \equiv \Delta q^f (x, Q^2) - \Delta \bar{q}^f
  (x, Q^2)
\end{align}
along with other flavor non-singlet helicity observables. First of
all, by analogy to the flavor singlet case, we need to define the
observables. Again we consider the diagrams in
\fig{f:helicity_TMD}. However, in the flavor non-singlet case we need
to {\sl subtract} from them the same diagrams with the quark particle
number flowing in the opposite direction in the quark loop. We obtain
the following expressions for flavor non-singlet helicity observables
(cf. Eqs.~\eqref{e:observables_S}):
\begin{subequations} \label{e:observables_NS}
  \begin{align} 
    g^{NS}_1 (x, Q^2) &= \frac{N_c}{2 \, \pi^2 \alpha_{EM}}
    \int\limits_{z_i}^1\frac{dz }{z^2 (1-z)} \, \int d x_{01}^2 \,
    \left[ \half \sum_{\lambda \sigma \sigma'} | \psi_{\lambda \sigma
        \sigma'}^T |^2_{(x_{01}^2 , z)} + \sum_{\sigma \sigma'}
      |\psi_{\sigma \sigma'}^L|^2_{(x_{01}^2 , z)} \right] G^{NS}
    (x_{01}^2 , z) ,
  \\
  \Delta q^{NS} (x, Q^2) &= \frac{N_c}{2 \pi^3} \int\limits_{z_i}^1
  \frac{dz}{z} \int\limits_{\tfrac{1}{z s}}^{\tfrac{1}{z Q^2}} \frac{d
    x_{01}^2}{x_{01}^2} \, G^{NS} (x_{01}^2 , z) ,
  \\
  g_{1L}^{NS} (x, k_T^2) &= \frac{8 \, N_c}{(2\pi)^6}
  \int\limits_{z_i}^1 \frac{dz}{z} \int d^2 x_{01} \, d^2 x_{0'1} \:
  e^{-i \ul{k} \cdot (\ul{x}_{01} - \ul{x}_{0' 1})} \:
  \frac{\ul{x}_{01} \cdot \ul{x}_{0'1}}{x_{01}^2 x_{0'1}^2} \, G^{NS}
  (x_{01}^2 , z) . \label{g1LNS}
  \end{align}
\end{subequations}
Here we do not sum over flavors. Hence our expressions
\eqref{e:observables_NS} should be understood as containing the
contribution of one particular quark flavor. Since quark masses can be
neglected in our DLA approximation, the small-$x$ asymptotics of all
helicity observables in \eqref{e:observables_S} and
\eqref{e:observables_NS} (flavor singlet and non-singlet) is
flavor-independent.  Flavor dependence may reside in the initial
conditions, but the asymptotic $x$ dependence (the intercept
$\alpha_h$ with $\Delta q \sim \left( \tfrac{1}{x}
\right)^{\alpha_h}$) should be independent of flavor.

Eqs.~\eqref{e:observables_NS} contain the flavor non-singlet polarized
dipole amplitude defined by (cf. Eqs.~\eqref{e:Gdef})
\begin{subequations}\label{e:GdefNS}
  \begin{align} 
    \label{e:Gdef1NS} 
    & G^{NS}_{10} (z) \equiv \frac{1}{2 N_c} \llangle \tr \left[
      V_{\ul 0} V_{\ul 1}^{pol \, \dagger} \right] - \tr \left[V_{\ul
        1}^{pol} V_{\ul 0}^\dagger \right] \rrangle (z) = G^{NS}
    (\ul{x}_1 , \ul{x}_0 , z) = G^{NS} (\ul{x}_{10} , \ul{b} , z) ,
   \\ \label{e:Gdef2NS}
   & G^{NS} (x_{01}^2 , z) \equiv \int d^2 b \: G^{NS}_{10} (z) .
 \end{align}
\end{subequations}
In Eqs.~\eqref{e:observables_NS} we assume that the leading
high-energy contribution to $G^{NS}_{10} (z)$ is real, such that
$G^{NS}_{10} (z) = G^{NS \, *}_{10} (z)$.

To determine the initial conditions $G_{10}^{NS \, (0)} (z)$ for the
flavor non-singlet helicity evolution we use Eqs.~\eqref{e:Gdef1NS}
and \eqref{VVsigma} to write
\begin{align} 
  \label{e:GdefNS3}
  G^{NS}_{10} (z) = - \frac{z s}{2} \left( \frac{d \sigma}{d^2 b}
    \Big[ q_{\ul{0}}^{unp} , \Delta\bar{q}_{\ul{1}} , z s \Big] -
    \frac{d \sigma}{d^2 b} \Big[ {\bar q}_{\ul{0}}^{unp} , \Delta
    q_{\ul{1}} , z s \Big] \right).
\end{align} 
Employing \eq{e:GdefNS3} we obtain for a single-quark target
\begin{subequations} \label{e:ICNS}
  \begin{align} 
  \label{e:ICNS1} 
  G_{10}^{NS \, (0)} &= \frac{\alpha_s^2 (C_F)^2}{N_c}
  \frac{1}{x_1^2}, %
  \\ \label{e:ICNS2}
  G^{NS \, (0)} (x_{10}^2 , z) &= \frac{\alpha_s^2 (C_F)^2}{N_c} \pi
  \ln\frac{z s}{\Lambda^2} .
 \end{align}
\end{subequations}
In arriving at Eqs.~\eqref{e:ICNS} we had to subtract the contribution
of the diagrams in the bottom row of \fig{f:init_conds} out of the
contribution of the diagrams in the top row of the same figure. The
last two diagrams in each row canceled, and the final answer in
Eqs.~\eqref{e:ICNS} is given solely by the upper left graph in
\fig{f:init_conds} with the $t$-channel quarks exchange. This makes
clear physical sense: only quarks can transfer flavor information from
the target to projectile. Therefore, quark exchange in the $t$-channel
is necessary for the flavor non-singlet observables.

Once again, the Born-level results \eqref{e:ICNS} can be ``dressed''
by GM/MV multiple rescatterings
\begin{align}
  G_{10}^{NS \, (0)} &= \frac{\alpha_s^2 (C_F)^2}{N_c} \frac{1}{x_1^2}
  \, \exp\left[-\tfrac{1}{4} \, x_{10}^2 \, Q_s^2 (b) \,
    \ln\tfrac{1}{x_{01} \Lambda} \right] . \label{e:ICNS3}
\end{align}
Either Eqs.~\eqref{e:ICNS} or \eqref{e:ICNS3} can be used as the
initial conditions for flavor non-singlet helicity evolution: if one
wants to keep saturation effects in the initial conditions one should
use \eqref{e:ICNS3}, otherwise one should use \eqref{e:ICNS}.

\subsection{Flavor Non-Singlet Helicity Evolution at Small $x$}

\label{FNSevol}

To construct flavor non-singlet helicity evolution equations, let us
consider one step of small-$x$ evolution. Looking at the diagrams in
Fig.~\ref{f:evol_step_latex}, we see immediately that the radiation of
soft polarized gluons carries polarization information but not flavor
and therefore does not contribute to the flavor non-singlet
distribution. Hence the first two diagrams on the right of the
equation illustrated in \fig{f:evol_step_latex} do not contribute.

\begin{figure}[hbt]
 \centering
 \includegraphics[width=\textwidth]{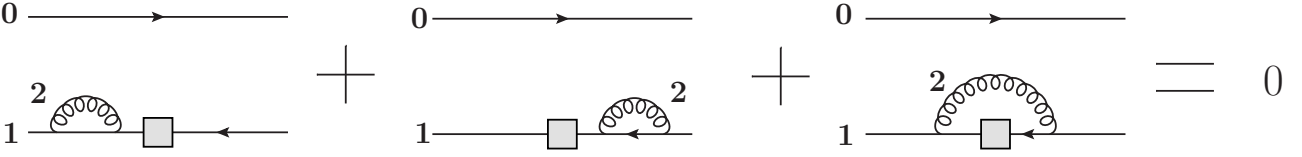}
 \caption{Real-virtual cancellations of soft unpolarized gluon
   emissions in the flavor non-singlet case (see \cite{Chen:1995pa}).
   In the strict DLA limit, only the polarized (anti)quark line
   interacts (as indicated by the absence of a shock-wave (blue
   rectangle)), and the sum of the real and virtual BFKL-like diagrams
   is zero.}
\label{f:real_virt_cancel_1}
\end{figure}

Next we consider the unpolarized-gluon emission diagrams in the bottom
row of \fig{f:evol_step_latex}. As we saw in deriving
Eqs.~\eqref{e:ICNS}, in the flavor non-singlet case the unpolarized
quark in the dipole amplitude does not interact in the initial
conditions. Hence, if the shock wave in \fig{f:evol_step_latex}
represented the initial conditions only, the diagrams in the bottom
row of that figure should cancel, as illustrated in
\fig{f:real_virt_cancel_1}. The same is true for the strict DLA
evolution case: the cancellation of \fig{f:real_virt_cancel_1} is true
at each step of the evolution, since it is true in the initial
conditions.

If one wants to go beyond the strict DLA limit and use the
saturation-enhanced initial conditions \eqref{e:ICNS3}, then soft
gluon in the right-most diagram of \fig{f:real_virt_cancel_1} would
interact with the target, and the cancellation would no longer be
valid. However, since the dipole size $x_{10}$ dependence in
\eq{e:ICNS3} is modified as compared to the Born-level
Eqs.~\eqref{e:ICNS} (for the terms containing multiple rescatterings),
the corresponding evolution is not going to be DLA, and would be
simply LLA BK/JIMWLK evolution in the dipole 01. In the large-$N_c$
limit such LLA evolution in dipole 01 is included in the right-most
diagram in the top row of \fig{f:evol_step_latex}. One also has to
include this LLA evolution into the initial conditions for the DLA
evolution \cite{Itakura:2003jp}. Let us stress one more time that such
corrections are beyond the strict DLA limit.

We are left only with the right-most diagram in the top row of
\fig{f:evol_step_latex} as contributing to the small-$x$ evolution of
the polarized dipole operator at small-$x$. Iteration of this diagram
would give us a simple ladder with quarks in the $t$-channel and with
gluon rungs (cf. \cite{Itakura:2003jp}).

\begin{figure}[hbt]
 \centering
 \includegraphics[width=0.95 \textwidth]{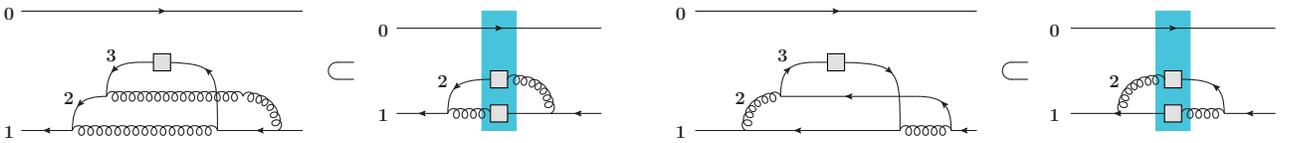}
 \caption{Some of the non-ladder diagrams contributing to the
   small-$x$ flavor non-singlet helicity evolution.}
\label{f:non-singl-subl_Nc}
\end{figure}

However, one should be careful here. We have never shown that the
diagrams in \fig{f:evol_step_latex} present all the possibilities for
one step of the flavor non-singlet helicity evolution. (Actually,
diagrams in \fig{f:evol_step_latex} illustrate the evolution in the
flavor singlet case, as derived in \cite{Kovchegov:2015pbl}.)  In
fact, for a step of the flavor non-singlet evolution, the diagrams in
\fig{f:evol_step_latex} do not exhaust all the possibilities. Some of
the flavor-singlet evolution diagrams not obtainable by evolution
equations employing multiple applications of the diagrams in
\fig{f:evol_step_latex} (or, more precisely, in Figs.~11 and 12 of
\cite{Kovchegov:2015pbl}) are shown in
\fig{f:non-singl-subl_Nc}. These are non-ladder diagrams which do not
contribute to the flavor singlet case, and which are not included in
\eq{e:opevol1}. The diagrams in \fig{f:non-singl-subl_Nc} are DLA and
should appear after two steps of flavor non-singlet helicity
evolution.

It appears that in order to include the diagrams from
\fig{f:non-singl-subl_Nc} into the operatorial helicity evolution
equations akin to \eqref{e:opevol1} one may need to define a polarized
``Wilson line'' that starts as a gluon/quark on one side of the shock
wave, and becomes a quark/gluon on the other side; examples of how
such evolution could play out are shown in
Fig.~\ref{f:non-singl-subl_Nc}.  These features will likely complicate
the formalism; indeed, since such ``identity-changing'' Wilson lines
occur in pairs, the evolution equations may be nonlinear. Luckily the
diagrams in question are subleading in $N_c$ and do not need to be
considered in the large-$N_c$ limit.

\begin{figure}[hbt]
 \centering
 \includegraphics[width=0.7 \textwidth]{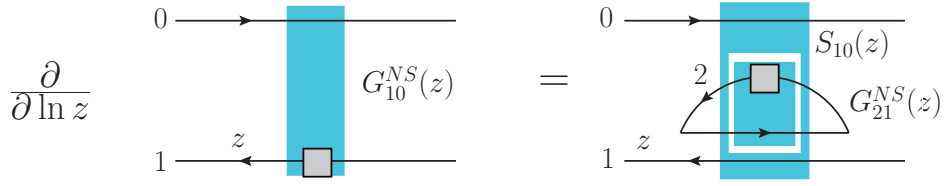}
 \caption{Large-$N_c$ evolution equation for the flavor non-singlet
   polarized dipole amplitude.  For simplicity, the initial conditions
   $G_{10}^{NS \, (0)}$ are not shown.}
\label{f:Nonsinglet_Large-N}
\end{figure}

We proceed by imposing the large-$N_c$ limit on the flavor non-singlet
helicity evolution. The resulting evolution equation receives
contributions only from the quark ladder (third diagram on the
right-hand side of Fig~\ref{f:evol_step_latex}); it is mathematically
almost identical to the familiar Reggeon evolution equation known in
the small-$x$ literature \cite{Itakura:2003jp}
\begin{align}
  \label{eq:NSevol33}
  G_{10}^{NS} (z) &= G_{10}^{NS \, (0)} (z) + \frac{\alpha_s
    N_c}{4\pi} \int\limits_{\tfrac{\Lambda^2}{s}}^z \frac{d z'}{z'}
  \int\limits_{\tfrac{1}{z' s}}^{x_{10}^2 \tfrac{z}{z'}} \frac{d
    x_{21}^2}{x_{21}^2} \, S_{10} (z') \, G_{21}^{NS} (z').
\end{align}
This equation is illustrated diagrammatically in
\fig{f:Nonsinglet_Large-N}.

In \eq{eq:NSevol33} we again include the non-linear LLA evolution
effects by keeping $S_{10} (z')$, to be found from the BK/JIMWLK
evolution, in the integrand. Note that the initial conditions can also
include BK/JIMWLK-evolved $S$-matrix in place of the exponential in
\eq{e:ICNS3}.


\subsection{Solution of flavor non-singlet helicity evolution
  equations at large $N_c$}

\label{FNSsol}

In the strict DLA limit we put $S=1$ everywhere, and \eq{eq:NSevol33}
becomes
\begin{align}
  \label{eq:NSevol4}
  G_{10}^{NS} (z) &= G_{10}^{NS \, (0)} (z) + \frac{\alpha_s
    N_c}{4\pi} \int\limits_{\tfrac{\Lambda^2}{s}}^z \frac{d z'}{z'}
  \int\limits_{\tfrac{1}{z' s}}^{x_{10}^2 \tfrac{z}{z'}} \frac{d
    x_{21}^2}{x_{21}^2} \, G_{21}^{NS} (z').
\end{align}
Integrating over the impact parameters yields
\begin{align} 
  \label{e:NSevol3}
  G^{NS} (x_{10}^2 , z) = G^{NS \, (0)} (x_{10}^2 , z) &+
  \frac{\alpha_s N_c}{4 \pi} \int\limits_{\Lambda^2 /s}^z \frac{d
    z'}{z'} \, \int\limits_{1/z' s}^{x_{10}^2 \tfrac{z}{z'}} \frac{d
    x_{21}^2}{x_{21}^2} G^{NS} (x_{21}^2 , z') .
\end{align}
The Reggeon evolution equation \eqref{e:NSevol3} can be solved by the
usual method of the Laplace-Mellin transform
\begin{align}
  \begin{aligned}
    G^{NS} (x_{10}^2 , z) &= \int\frac{d\omega}{2\pi i} e^{\omega
      \eta} \int\frac{d\lambda}{2\pi i} e^{\lambda s_{10}} \:
    G_{\omega \lambda}^{NS}  ,
    \\
    G^{NS}_{\omega \lambda} &= \int\limits_0^\infty d (\eta - s_{10})
    \, e^{-\lambda (\eta - s_{10})} \int\limits_0^\infty d \eta \,
    e^{- \omega \eta} \, G^{NS}(x_{10}^2 , z) ,
  \end{aligned}
\end{align}
where the natural variables for the transform are
\begin{align}
\begin{aligned}
  \eta &\equiv \ln\tfrac{z s}{\Lambda^2} > 0& \hspace{2cm}
  \eta' &\equiv \ln\tfrac{z' s}{\Lambda^2} > 0 \\
  s_{10} &\equiv \ln \left( \frac{1}{x_{10}^2 \Lambda^2} \right) <
  \eta& s_{21} &\equiv \ln \left( \frac{1}{x_{21}^2 \Lambda^2} \right)
  < \eta' .
\end{aligned}
\end{align}
In terms of these variables, the flavor non-singlet evolution equation
is
\begin{align}
  \label{eq:NS1}
  G^{NS} (s_{10} , \eta) = G^{NS \, (0)} (s_{10} , \eta) &+
  \frac{\alpha_s N_c}{4 \pi} \int\limits_{0}^\eta d \eta' \,
  \int\limits_{s_{10} - \eta + \eta'}^{\eta'} d s_{21} G^{NS} (s_{21}
  , \eta') .
\end{align}
In Mellin space, the evolution equation is solved algebraically:
\begin{align}
  G^{NS}_{\omega \lambda} = \frac{1}{1-\left(\tfrac{\alpha_s
        N_c}{4\pi}\right) \tfrac{1}{\omega \lambda}} G_{\omega
    \lambda}^{NS \, (0)} = \frac{\alpha_s^2 (C_F)^2 \pi}{N_c \,
    \omega} \left(\frac{1}{\omega \lambda - \tfrac{\alpha_s
        N_c}{4\pi}}\right) ,
\end{align}
where the flavor non-singlet initial condition comes from
\eq{e:ICNS2}.  The large-$(z s)$ asymptotics, evaluated in the saddle
point approximation, are given by
\begin{align} 
  \label{e:Reggeon}
  G^{NS} (x_{10}^2 , z) \propto ( z s )^{\alpha_h^{NS}} \hspace{1cm}
  \mathrm{with} \hspace{1cm} \alpha_h^{NS} = {\sqrt{\tfrac{\alpha_s
        N_c}{\pi}}} ,
\end{align}
such that 
\begin{align}
  g_1^{NS} (x, Q^2) \sim \Delta q^{NS} (x, Q^2) \sim g_{1L}^{NS} (x,
  k_T^2) \sim \left( \frac{1}{x} \right)^{\alpha_{h}^{NS}} \approx
  \left( \frac{1}{x} \right)^{{\sqrt{\tfrac{\alpha_s N_c}{\pi}}}}.
\end{align}
Note again that the intercept is flavor-independent at this leading
order obtained by the DLA resummation. 

This flavor non-singlet intercept $\alpha_h^{NS}$, calculated from
solving our large-$N_c$ non-singlet helicity evolution equation
\eqref{eq:NSevol4}, agrees exactly with the (large-$N_c$ limit of the)
result of BER's calculation \cite{Bartels:1995iu} using the method of
infrared evolution equations.  Progress in incorporating nonlinear
multiple scattering corrections to the Reggeon-like evolution
equations like \eqref{eq:NSevol33} has been made in the context of
baryon number transport at small $x$ \cite{Itakura:2003jp}.


\section{Conclusions}
\label{sec:Conclusions}

In this paper we have considered small-$x$ asymptotics of the flavor
singlet and non-singlet helicity observables. We have defined the
relations between the helicity TMD's, PDF's and $g_1$ structure
functions to the polarized dipole operators in both flavor singlet and
non-singlet cases. The resulting difference in the polarized dipole
operators can be seen in Eqs.~\eqref{e:Gdef1} and
\eqref{e:Gdef1NS}. We have evaluated the polarized dipole amplitude in
the MV model/GM approximation obtaining the initial conditions for the
small-$x$ evolution. We have then re-constructed evolution equations
for the polarized dipole amplitude in the flavor singlet case
originally derived in \cite{Kovchegov:2015pbl}, filling in the
important intermediate steps not presented in
\cite{Kovchegov:2015pbl}.

The solution of the large-$N_c$ flavor singlet evolution equations,
presented in \cite{Letter}, leads to the intercept in \eq{hel_int},
which is about 2/3 of the flavor singlet intercept obtained by BER in
\cite{Bartels:1996wc}. Our calculation satisfies all of the same
cross-checks as BER (with the exception of the NNLO anomalous
dimension for polarized DGLAP which we did not verify due to
complexity of the calculation in our approach). Our effort to
reproduce the calculation of BER working in Feynman gauge used in
\cite{Bartels:1996wc} is presented in Appendix~\ref{sec:BERcheck}. At
the moment it appears that BER might be missing parts of the DLA
contributions of diagrams $B$, $C$, $D$, $E$ and $I$ from
\fig{Feyn_gauge23} in their calculation.

It is possible that, by omitting the DLA contributions discussed in
our Appendix~\ref{sec:BERcheck}, BER effectively restricted their
analysis to the leading-twist evolution only (see Eq. (4.1) in
\cite{Bartels:1996wc}), or, at least discarded a subset of
higher-twist terms. This assumption is consistent with the BER
formalism generating correct anomalous dimensions for polarized DGLAP
evolution, presently verified up to (and including) NNLO
\cite{Moch:2014sna}. The disagreement between BER and our intercept
may then be attributed to the fact that our evolution is all-twist,
due to the terms like $\theta ( \tfrac{\mu^2}{Q^2} - x )$
which we include in our helicity evolution (see the discussion below
\eq{e:Kmod} where such terms were mentioned, and neglected, but only
in the DGLAP anomalous dimension calculation). (The discontinuous
nature of the theta-function terms is probably a property of DLA and
is likely to be smoothed-out by higher-order corrections.) In the case
of unpolarized BFKL evolution, which is all-twist, it is known that
the exact all-twist intercept $\alpha_P -1 = \tfrac{4 \as N_c}{\pi}
\ln 2$ is about 30$\%$ smaller than the leading-twist contribution to
the intercept which yields $(\alpha_P -1)_{LT} = \tfrac{4 \as
  N_c}{\pi}$ (see the discussion on pp. 246-247 of
\cite{Kovchegov:2012mbw}). It is possible that something similar takes
place in the helicity evolution case at hand, accounting for the
difference between the leading-twist BER calculation
\cite{Bartels:1996wc} and our all-twist intercept \eqref{hel_int}.

For the flavor non-singlet helicity evolution we have derived the
large-$N_c$ evolution equation \eqref{eq:NSevol33}. The resulting
intercept \eqref{e:Reggeon} is in compete agreement with BER
\cite{Bartels:1995iu}.

To summarize the status of the leading-order calculations of various
intercepts mainly resulting from the DLA evolution, in
Table~\ref{table:Intercepts} we list the intercepts for flavor singlet
and non-singlet evolution for the unpolarized and helicity-dependent
observables. The intercepts for helicity evolution were obtained by us
and by BER in various approximations.

It is important to get a better understanding of the numerical
importance of these results for the small-$x$ contribution to the
quark spin of the proton,
\begin{align}
  \label{eq:quark_spin}
  S_q (Q^2) = \frac{1}{2} \, \int\limits_0^1 dx \, \Delta \Sigma (x,
  Q^2), 
\end{align}
with 
\begin{align}
  \label{eq:Sigma}
  \Delta \Sigma (x, Q^2) = \left[ \Delta u + \Delta {\bar u} + \Delta
    d + \Delta {\bar d} + \ldots \right] \! (x, Q^2).
\end{align}
A detailed analysis of the impact of our flavor singlet intercept on
$\Delta \Sigma (x, Q^2)$ at small $x$ is carried out in
\cite{Letter}. Clearly, large intercepts may potentially lead to a
divergent integral in \eq{eq:quark_spin}, and would require
higher-order corrections or saturation effects at small $x$ to make
the integral finite.

\begin{table}[htb]
 \begin{tabular}{ |l|l|c|c|c|c| }
  \hline
           &        &               & $Q^2 = 3$~GeV$^2$     & $Q^2 = 10$~GeV$^2$  &  $Q^2 = 87$~GeV$^2$ \\
 Observable        & Evolution               & Intercept & $\alpha_s = 0.343$  & $\alpha_s = 0.249$ & $\alpha_s = 0.18$ \\ \hline \hline
   Unpolarized flavor singlet &  LO BFKL Pomeron        & $1 + \tfrac{\alpha_s N_c}{\pi} \, 4 \ln 2$  & $1.908$ & $1.659$ & $1.477$ \\
   structure function $F_2$ & & & & & \\ \hline
     Unpolarized flavor non-singlet &  Reggeon        & $\sqrt{\tfrac{2 \, \alpha_s C_F}{\pi}}$ & $0.540$ & $0.460$ & $0.391$ \\
   structure function $F_2$ & & & & & \\
    \hline \hline
     Flavor singlet &   us (Pure Glue, Large-$N_c$)  & $2.31 \, \sqrt{\tfrac{\alpha_s N_c}{2\pi}}$ & $0.936$ & $0.797$ & $0.678$ \\
   structure function $g_1^S$ &  BER (Pure Glue)  & $3.66 \, \sqrt{\tfrac{\alpha_s N_c}{2\pi}}$ & $1.481$ & $1.262$ & $1.073$ \\ 
   & BER $(N_f = 4)$ & $3.45 \, \sqrt{\tfrac{\alpha_s N_c}{2\pi}}$ & $1.400$ & $1.190$ & $1.011$ \\ \hline
       Flavor non-singlet &   BER and us (Large-$N_c$)  & $\sqrt{\tfrac{\alpha_s N_c}{\pi}}$ & $0.572$ & $0.488$ & $0.415$ \\
   structure function $g_1^{NS}$ &  & & & &  \\ \hline
 \end{tabular}
 \caption{Comparison of the intercepts $\alpha$ leading to helicity PDF's which scale as $\Delta q_f (x, Q^2) \propto (\tfrac{1}{x})^\alpha$ in  the high-energy / small-$x$ asymptotics.  The LO BFKL Pomeron which sets the small-$x$ asymptotics of unpolarized PDF's is shown for comparison, along with the the LO intercept of the perturbative QCD Reggeon.  Unless otherwise specified, the quoted intercepts are taken at finite $N_c$.}
 \label{table:Intercepts}
\end{table}

To see which of the small-$x$ helicity intercepts give a finite
integral in \eq{eq:quark_spin}, we compute their numerical values in
Table~\ref{table:Intercepts} for $N_c = 3$ and $\alpha_s$ set by the
one-loop running coupling expression
\begin{align}
  \alpha_s (Q^2) = \frac{4\pi}{(11 - \tfrac{2}{3} N_f) \ln
    \tfrac{Q^2}{\Lambda^2}}
\end{align}
with $\Lambda = 0.192$~GeV and $N_f = 3$ for purposes of the
scale-setting. (Since this is a rough estimate, and includes a
pure-glue and fixed-$N_f$ numerical estimates of the intercept, we do
not change our $N_f$ with $Q^2$ for simplicity.) For comparison, we
have included the leading-order (LO) BFKL intercept
\cite{Kuraev:1977fs,Balitsky:1978ic} along with the intercept for the
perturbative QCD Reggeon \cite{Kirschner:1983di, Kirschner:1985cb,
  Kirschner:1994vc, Kirschner:1994rq, Griffiths:1999dj,
  Itakura:2003jp}. We see that, for a wide range of $Q^2$, the BER
results generate small-$x$ intercepts which are greater than $1$ and
hence non-integrable.  Our result, on the other hand, generally yields
an integrable singularity at $x \rightarrow 0$. Indeed this only means
that our result would not require higher-order or saturation
corrections to give a finite integral in \eq{eq:quark_spin}.

Note that a strong 't Hooft coupling calculation \cite{Hatta:2009ra}
in the framework of the anti-de Sitter/conformal field theory
(AdS/CFT) correspondence appears to indicate that in ${\cal N} =4$
super--Yang--Mills theory the flavor non-singlet intercept is smaller
than one for all couplings, with the flavor singlet contribution being
suppressed at large coupling. If this conclusion applies to QCD, this
may indicate that higher-order correction would not allow any of the
perturbative intercepts found in this work ($\alpha_h$ or
$\alpha_h^{NS}$) to exceed unity.

In addition, higher order corrections are needed to obtain a more
reliable comparison with the experimental data.  Future work on the
subject would include solving the flavor singlet evolution equations
derived in \cite{Kovchegov:2015pbl} for the large-$N_c \, \& N_f$
limit, which includes quarks. Including running coupling corrections
using the Brodsky--Lepage--Mackenzie (BLM) \cite{BLM} scheme along the
lines of
\cite{Balitsky:2006wa,Kuokkanen:2011je,Kovchegov:2006vj,Kovchegov:2006wf}
for unpolarized evolution (see also
\cite{Ermolaev:1999jx,Ermolaev:2000sg,Ermolaev:2003zx} for other
methods used for helicity evolution) would be a natural next step
ultimately leading to a detailed comparison to the experimental
longitudinal spin data at small $x$, complementing the existing
approaches
\cite{Soffer:1996ft,Kiyo:1996si,Bluemlein:2002be,Ermolaev:2009cq,deFlorian:2009vb,deFlorian:2014yva,Aschenauer:2015ata}.


\section*{Acknowledgments}

YK is grateful to Boris Ermolaev for a discussion of BER calculation,
to Jochen Bartels for correspondence, and to Ian Balitsky for an
informative discussion. This material is based upon work supported by
the U.S. Department of Energy, Office of Science, Office of Nuclear
Physics under Award Number DE-SC0004286 (YK), within the framework of
the TMD Topical Collaboration (DP), and DOE Contract No. DE-SC0012704
(MS).  DP also received support from the RIKEN BNL Research Center. MS
received additional support from an EIC program development fund from
BNL and from the U.S. Department of Energy, Office of Science under
the DOE Early Career Program. \\


\appendix
\section{Taking the large-$N_c$ limit of helicity evolution}
\label{sec:largeN}

The large-$N_c$ limit means that different dipoles do not 'talk' to
each other in the process of evolution and interaction with the
target. However, when we write a gluon line as a double
(quark-antiquark) line, it is a statement only about color factors:
this does not mean that all the other dynamical factors associated
with the gluon dynamics also split into those for quark and
antiquark. Namely, the $G \to G G$ splitting wave function is not, in
general, equal to the sum of $q \to q G$ and ${\bar q} \to {\bar q} G$
wave functions. Confusion may arise because in the eikonal limit the
$G \to G G$ splitting is, in fact, a sum of $q \to q G$ and ${\bar q}
\to {\bar q} G$ wave functions.

To demonstrate this in our case let us start with the evolution
equation for the adjoint dipole (Eq.~(62) from
\cite{Kovchegov:2015pbl}) keeping flavor-singlet evolution in mind
\begin{align}\label{Gevol3all}
  & \frac{1}{N_c^2 -1} \, \left\langle \!\! \left\langle \mbox{Tr}
      \left[ U_{\ul{0}} \, U_{\ul{1}}^{pol \, \dagger} \right]
    \right\rangle \!\! \right\rangle (z) = \frac{1}{N_c^2 -1} \,
  \left\langle \!\! \left\langle \mbox{Tr} \left[ U_{\ul{0}} \,
        U_{\ul{1}}^{pol \, \dagger} \right] \right\rangle \!\!
  \right\rangle_0 (z) + \frac{\as}{2 \pi^2} \int\limits_{\Lambda^2 /
    s}^z \frac{d z'}{z'} \, \int \frac{d^2 x_{2}}{x_{21}^2} \,
  \theta(x_{21}^2 - \tfrac{1}{z' s}) \notag \\ & \times \left\{ \theta
    (x_{10} - x_{21}) \, \frac{4}{N_c^2 -1} \left\langle \!\!
      \left\langle \mbox{Tr} \left[ T^b \, U_{\ul{0}} \, T^a \,
          U_{\ul{1}}^{\dagger} \right] \, \left( U^{pol}_{\ul{2}}
        \right)^{ba} \right\rangle \!\!  \right\rangle (z')
  \right. \notag \\ & - \theta (x_{10}^2 z - x_{21}^2 z') \,
  \frac{N_f}{N_c^2-1} \left\langle \!\! \left\langle \mbox{tr} \left[
        t^b \, V_{\ul{1}} \, t^a \, V_{\ul{2}}^{pol \, \dagger}
      \right] \, U^{ba}_{\ul{0}} + \mbox{tr} \left[ t^b \,
        V^{pol}_{\ul{2}} \, t^a \, V_{\ul{1}}^{\dagger} \right] \,
      U^{ba}_{\ul{0}} \right\rangle \!\! \right\rangle (z') \notag \\
  & \left. + \theta (x_{10} - x_{21}) \frac{2}{N_c^2 -1} \left[
      \left\langle \!\!  \left\langle \mbox{Tr} \left[ T^b \,
            U_{\ul{0}} \, T^a \, U_{\ul{1}}^{pol \, \dagger} \right]
          \, U^{ba}_{\ul{2}} \right\rangle \!\!  \right\rangle (z') -
      N_c \left\langle \!\! \left\langle \mbox{Tr} \left[ U_{\ul{0}}
            \, U_{\ul{1}}^{pol \, \dagger} \right] \right\rangle \!\!
      \right\rangle (z') \right] \right\}.
\end{align}
Here the $U$'s are adjoint Wilson lines. Concentrating on the term
responsible for the emission of the polarization-carrying soft gluon
we write \eq{Gevol3all} as
\begin{align}\label{Gevol3}
  & \frac{1}{N_c^2 -1} \, \left\langle \!\! \left\langle \mbox{Tr}
      \left[ U_{\ul{0}} \, U_{\ul{1}}^{pol \, \dagger} \right]
    \right\rangle \!\! \right\rangle (z) = \frac{1}{N_c^2 -1} \,
  \left\langle \!\! \left\langle \mbox{Tr} \left[ U_{\ul{0}} \,
        U_{\ul{1}}^{pol \, \dagger} \right] \right\rangle \!\!
  \right\rangle_0 (z) + \frac{\as}{2 \pi}
  \int\limits_{\tfrac{1}{x_{10}^2 s}}^z \frac{d z'}{z'} \,
  \int\limits_{\tfrac{1}{z' s}}^{x_{01}^2} \frac{d x^2_{21}}{x_{21}^2}
  \notag \\ & \times \left\{ \frac{4}{N_c^2 -1} \left\langle \!\!
      \left\langle \mbox{Tr} \left[ T^b \, U_{\ul{0}} \, T^a \,
          U_{\ul{1}}^{\dagger} \right] \, \left( U^{pol}_{\ul{2}}
        \right)^{ba} \right\rangle \!\!  \right\rangle (z') + \ldots
  \right\}.
\end{align}
Our goal here is to properly take the large-$N_c$ limit of
\eqref{Gevol3}, writing the answer in terms of the fundamental dipole
operators. To do so we remember that
\begin{align}
  \label{Uunp}
  U^{ab} = 2 \, \mbox{tr} \left[ t^b V^{\dagger} t^a V \right],
\end{align}
as expected for Wilson lines.

To write a similar expression for the ``polarized adjoint Wilson
line'' operator, in the large-$N_c$ limit one can think of it as a
Wilson line with one insertion of a non-eikonal vertex (due to a
spin-dependent gluon exchange). Let us model this non-eikonal vertex
as a (possibly transverse) derivative acting on the true (unpolarized)
Wilson line (see e.g. \cite{Altinoluk:2014oxa}), that is, we write
\begin{align}
  \left( U^{pol}\right)^{ab} \propto \pd U^{ab} = \pd \left\{ 2 \,
    \mbox{tr} \left[ t^b V^{\dagger} t^a V \right] \right\} = 2 \,
  \mbox{tr} \left[ t^b (\pd V^{\dagger}) t^a V \right] + 2 \,
  \mbox{tr} \left[ t^b V^{\dagger} t^a (\pd V) \right] \notag \\
  \propto 2 \, \mbox{tr} \left[ t^b V^{pol \, \dagger} t^a V \right] +
  2 \, \mbox{tr} \left[ t^b V^{\dagger} t^a V^{pol} \right].
\end{align}
In the last step we identified $\pd V \to V^{pol}$. Using the
resulting relation
\begin{align}
  \label{Upol}
  \left( U^{pol}\right)^{ab} = 2 \, \mbox{tr} \left[ t^b V^{pol \,
      \dagger} t^a V \right] + 2 \, \mbox{tr} \left[ t^b V^{\dagger}
    t^a V^{pol} \right]
\end{align}
along with \eq{Uunp} to simplify the adjoint polarized dipole operator
on the left-hand side of \eq{Gevol3} we obtain
\begin{align}
  \mbox{Tr} \left[ U_{\ul{0}} \, U_{\ul{1}}^{pol \, \dagger} \right] =
  U^{ba}_{\ul{0}} \, \left( U_{\ul{1}}^{pol \, \dagger} \right)^{ab} =
  U^{ba}_{\ul{0}} \, \left( U_{\ul{1}}^{pol} \right)^{ba} = 4 \,
  \mbox{tr} \left[ t^a V_0^{\dagger} t^b V_0 \right] \, \left(
    \mbox{tr} \left[ t^a V_1^{pol \, \dagger} t^b V_1 \right] +
    \mbox{tr} \left[ t^a V_1^{\dagger} t^b V_1^{pol} \right] \right)
  \notag \\ = 2 \, \mbox{tr} \left[V_0^{\dagger} t^b V_0 \, V_1^{pol
      \, \dagger} t^b V_1 \right] + 2 \, \mbox{tr} \left[V_0^{\dagger}
    t^b V_0 \, V_1^{\dagger} t^b V_1^{pol} \right] = \mbox{tr} \left[
    V_0 \, V_1^{pol \, \dagger} \right] \, \mbox{tr} \left[V_1 \,
    V_0^{\dagger} \right] + \mbox{tr} \left[ V_0 \, V_1^{\dagger}
  \right] \, \mbox{tr} \left[V^{pol}_1 \, V_0^{\dagger} \right] +
  \ldots , \label{UUVV}
\end{align}
where the ellipsis denote the subleading-$N_c$ terms. In arriving at
the end result in \eq{UUVV} we have applied the Fierz identity twice.

We conclude that
\begin{align}\label{UUleft}
  \frac{1}{N_c^2 -1} \, \left\langle \!\! \left\langle \mbox{Tr}
      \left[ U_{\ul{0}} \, U_{\ul{1}}^{pol \, \dagger} \right]
    \right\rangle \!\! \right\rangle (z) = 2 \, {G}_{10} (z) \, S_{01}
  (z)
\end{align}
in the large-$N_c$ limit. Note that the polarized dipole amplitude
${G}_{10} (z)$ is made out of the quark lines coming from the gluons
in the large-$N_c$ limit.

Now let us perform a similar analysis to the operator on the
right-hand side of \eq{Gevol3}:
\begin{align}
  & \mbox{Tr} \left[ T^b \, U_{\ul{0}} \, T^a \, U_{\ul{1}}^{\dagger}
  \right] \, \left( U^{pol}_{\ul{2}} \right)^{ba} = - f^{bcd} U_0^{de}
  f^{aeg} \, \left(U_1^\dagger \right)^{gc} \, 2 \, \left\{ \mbox{tr}
    \left[ t^a V_2^{pol \, \dagger} t^b V_2 \right] + \mbox{tr} \left[
      t^a V_2^{\dagger} t^b V_2^{pol} \right] \right\} \notag \\ & = -
  8 \, f^{bcd} \, f^{aeg} \, \mbox{tr} \left[ t^e V_0^{\dagger} t^d
    V_0 \right] \, \mbox{tr} \left[ t^g V_1^{\dagger} t^c V_1 \right]
  \, \left\{ \mbox{tr} \left[ t^a V_2^{pol \, \dagger} t^b V_2 \right]
    + \mbox{tr} \left[ t^a V_2^{\dagger} t^b V_2^{pol} \right]
  \right\} \notag \\ & = 8 \, \mbox{tr} \left[ t^e V_0^{\dagger} [t^b,
    t^c] V_0 \right] \, \mbox{tr} \left[ [t^a, t^e] V_1^{\dagger} t^c
    V_1 \right] \, \left\{ \mbox{tr} \left[ t^a V_2^{pol \, \dagger}
      t^b V_2 \right] + \mbox{tr} \left[ t^a V_2^{\dagger} t^b
      V_2^{pol} \right] \right\}. \label{UUU1}
\end{align}
Let us concentrate on the first two traces: using Fierz identity
multiple times we write
\begin{align}
  & \mbox{tr} \left[ t^e V_0^{\dagger} [t^b, t^c] V_0 \right] \,
  \mbox{tr} \left[ [t^a, t^e] V_1^{\dagger} t^c V_1 \right] =
  \mbox{tr} \left[ t^e V_0^{\dagger} (t^b t^c - t^c t^b) V_0 \right]
  \, \mbox{tr} \left[ (t^a t^e - t^e t^a) V_1^{\dagger} t^c V_1
  \right] \notag \\ & = \frac{1}{2} \mbox{tr} \left[ V_0^{\dagger}
    (t^b t^c - t^c t^b) V_0 \, V_1^{\dagger} t^c V_1 t^a \right] -
  \frac{1}{2} \mbox{tr} \left[ V_0^{\dagger} (t^b t^c - t^c t^b) V_0
    \, t^a \, V_1^{\dagger} t^c V_1 \right] \notag \\ & = \frac{1}{4}
  \mbox{tr} \left[ V_0 V_1^{\dagger} \right] \, \mbox{tr} \left[ V_1
    t^a V_0^{\dagger} t^b \right] - \frac{1}{4} \mbox{tr} \left[ t^b
    V_0 V_1^{\dagger} \right] \, \mbox{tr} \left[ V_1 t^a
    V_0^{\dagger} \right] - \frac{1}{4} \mbox{tr} \left[ V_0 t^a
    V_1^{\dagger} \right] \, \mbox{tr} \left[ V_1 V_0^{\dagger} t^b
  \right] + \frac{1}{4} \mbox{tr} \left[ t^b V_0 t^a V_1^{\dagger}
  \right] \, \mbox{tr} \left[ V_1 V_0^{\dagger} \right]. \label{UUU2}
\end{align}
Substituting \eq{UUU2} into \eq{UUU1} and using Fierz identity two
more times we arrive at
\begin{align}
  & \mbox{Tr} \left[ T^b \, U_{\ul{0}} \, T^a \, U_{\ul{1}}^{\dagger}
  \right] \, \left( U^{pol}_{\ul{2}} \right)^{ba} = 2 \, \left\{
    \mbox{tr} \left[ t^a V_2^{pol \, \dagger} t^b V_2 \right] +
    \mbox{tr} \left[ t^a V_2^{\dagger} t^b V_2^{pol} \right] \right\}
  \notag \\ & \times \, \left\{ \mbox{tr} \left[ V_0 V_1^{\dagger}
    \right] \, \mbox{tr} \left[ V_1 t^a V_0^{\dagger} t^b \right] -
    \mbox{tr} \left[ t^b V_0 V_1^{\dagger} \right] \, \mbox{tr} \left[
      V_1 t^a V_0^{\dagger} \right] - \mbox{tr} \left[ V_0 t^a
      V_1^{\dagger} \right] \, \mbox{tr} \left[ V_1 V_0^{\dagger} t^b
    \right] + \mbox{tr} \left[ t^b V_0 t^a V_1^{\dagger} \right] \,
    \mbox{tr} \left[ V_1 V_0^{\dagger} \right] \right\} \notag \\ & =
  \frac{1}{2} \, \mbox{tr} \left[ V_0 V_1^{\dagger} \right] \,
  \mbox{tr} \left[ V^{pol}_2 V_0^{\dagger} \right] \, \mbox{tr} \left[
    V_1 V_2^{\dagger} \right] + \frac{1}{2} \, \mbox{tr} \left[ V_0
    V_1^{\dagger} \right] \, \mbox{tr} \left[ V_2 V_0^{\dagger}
  \right] \, \mbox{tr} \left[ V_1 V_2^{pol \, \dagger} \right] +
  \frac{1}{2} \, \mbox{tr} \left[ V_0 V_1^{\dagger} \right] \,
  \mbox{tr} \left[ V^{pol}_2 V_1^{\dagger} \right] \, \mbox{tr} \left[
    V_0 V_2^{\dagger} \right] \notag \\ & + \frac{1}{2} \, \mbox{tr}
  \left[ V_0 V_1^{\dagger} \right] \, \mbox{tr} \left[ V_2
    V_1^{\dagger} \right] \, \mbox{tr} \left[ V_0 V_2^{pol \, \dagger}
  \right] + \ldots, \label{UUU3}
\end{align}
where the ellipsis denote the $N_c$-suppressed term, which include
operators which are $N_c$-suppressed due to consisting of fewer than
three traces but not suppressed by explicit factors of $1/N_c$.

We thus conclude that in the large-$N_c$ limit
\begin{align}
  \label{UUUright}
  \frac{1}{N_c^2 -1} \left\langle \!\! \left\langle \mbox{Tr} \left[
        T^b \, U_{\ul{0}} \, T^a \, U_{\ul{1}}^{\dagger} \right] \,
      \left( U^{pol}_{\ul{2}} \right)^{ba} \right\rangle \!\!
  \right\rangle (z) = N_c \, S_{01} (z) \, \left[ S_{02} (z) \,
    {G}_{21} (z) + S_{21} (z) \, {\Gamma}_{02,21} (z) \right].
\end{align}
Substituting Eqs.~\eqref{UUleft} and \eqref{UUUright} into \eq{Gevol3}
we arrive at
\begin{align}
  \label{Gtilde_evol} {G}_{10} (z) = {G}^{(0)}_{10} (z) + \frac{\as \,
    N_c}{2 \, \pi} \int\limits_{\tfrac{1}{x_{10}^2 s}}^z \frac{d z'}{z'} \,
  \int\limits_{\tfrac{1}{z' s}}^{x_{01}^2} \frac{d x^2_{21}}{x_{21}^2} \left[
    2\, S_{02} (z') \, {G}_{21} (z') + 2 \, S_{21} (z') \,
    {\Gamma}_{02,21} (z') + \ldots \right],
\end{align}
in agreement with the first two terms in the integral on the right of
Eqs.~\eqref{e:largeNevol}.

Note that in arriving at \eq{Gtilde_evol} we have implicitly assumed that
\begin{align}\label{UUright}
  \frac{1}{N_c^2 -1} \, \left\langle \!\! \left\langle \mbox{Tr}
      \left[ U_{\ul{0}} \, U_{\ul{1}}^{pol \, \dagger} \right]
    \right\rangle \!\! \right\rangle_0 (z) = 2 \, {G}^{(0)}_{10} (z)
  \, S_{01} (z)
\end{align}
with a fully (LLA) evolved $S_{01} (z)$. Hence \eq{Gtilde_evol} can be
thought of as helicity evolution in the background of the unpolarized
LLA evolution.


\section{Reproducing BER}
\label{sec:BERcheck}

\begin{figure}[t]
\begin{center}
\includegraphics[width=0.95 \textwidth]{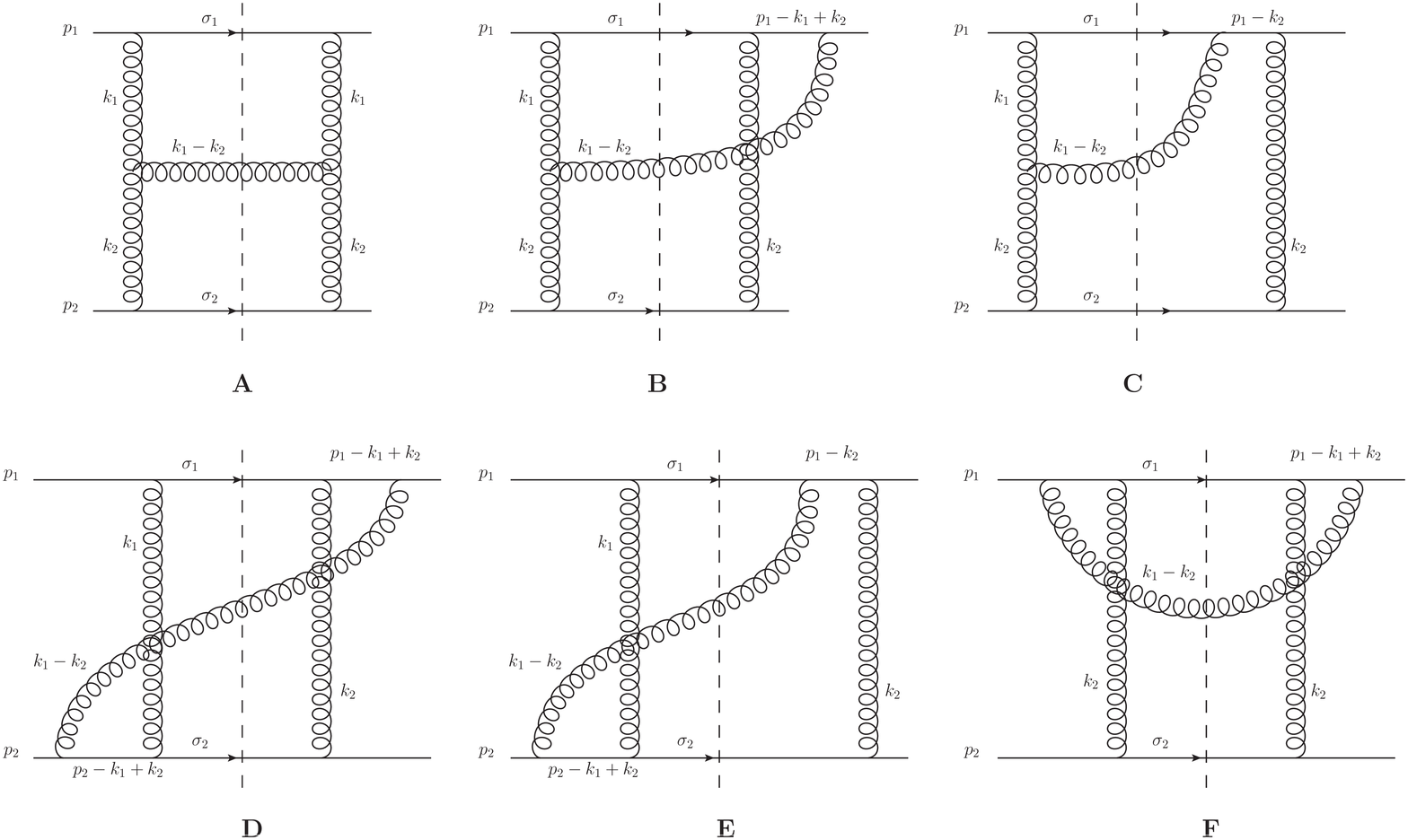} \\~~\\
\includegraphics[width=0.95 \textwidth]{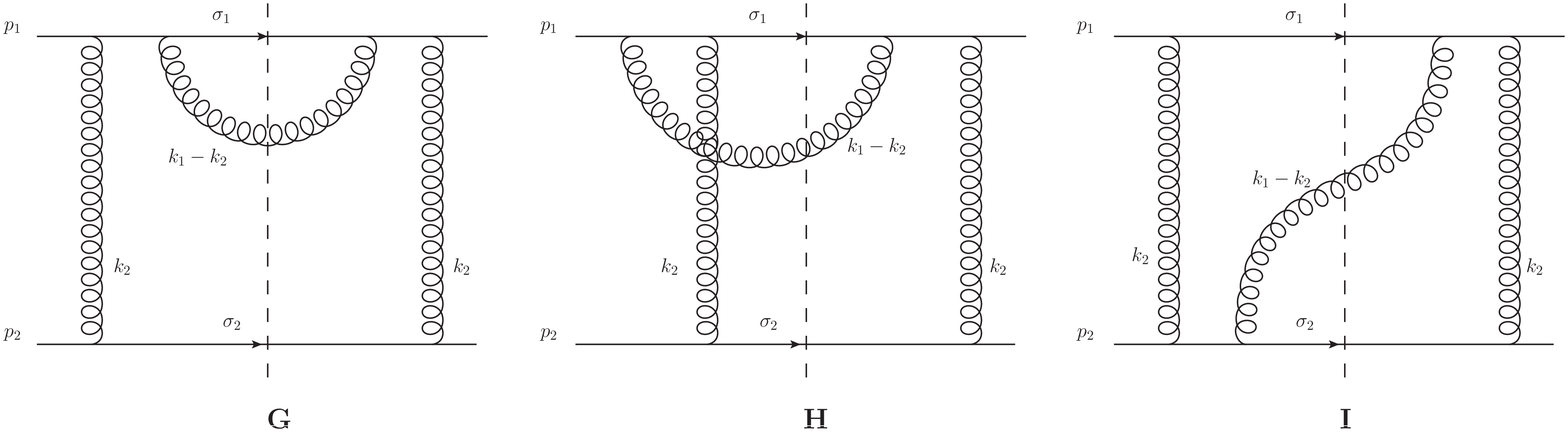}
\caption{Diagrams representing one-loop DLA helicity evolution
  corrections to the Born-level graphs in Feynman gauge. Dashed lines
  denote the final state cuts. Only corrections with the extra gluon
  going through the final state cut are considered.}
\label{Feyn_gauge23}
\end{center}
\end{figure}

In order to establish a connection between our work and the paper by
BER \cite{Bartels:1996wc} we tried calculating the first small-$x$
evolution correction to the Born-level cross section mediated by the
gluon exchanges. We performed our calculation in the Feynman gauge,
just like the authors of \cite{Bartels:1996wc} did. (Note that our
evolution calculations here and in \cite{Kovchegov:2015pbl} were done
in the light-cone gauge.) The diagrams we analyzed are shown in
\fig{Feyn_gauge23}, where we concentrate on real gluon emissions
only. As usual in high-energy scattering we work in the eikonal limit
where
\begin{align}
  p_1^+, p_2^- \gg k_1^+, k_2^-, k_{1\perp}, k_{2 \perp} \gg k_1^-,
  k_2^+ .
\end{align}
For simplicity we put $p_1^- = 0 = p_2^+$ and $\ul{p}_1 = 0 =
\ul{p}_2$.

We are interested in the parts of the diagrams contributing to the
double-spin asymmetry. Keeping the $\sigma_1 \, \sigma_2$ terms only
and performing a direct calculation we obtain the following leading in
energy contributions to the amplitude squared:
\begin{subequations}\label{ampl}
\begin{align}
  & \langle |M|^2 \rangle \bigg|_A = 16 g^6 \, C_F \, \sigma_1 \, \sigma_2 \, \frac{s}{k_{1 \perp}^2 \, k_{2 \perp}^2}, \\
  & \langle |M|^2 \rangle \bigg|_B = 4 g^6 \, C_F \, \sigma_1 \, \sigma_2 \, s \, \frac{\ul{k}_2 \cdot (\ul{k}_1 - 2 \ul{k}_2)}{k_{1 \perp}^2 \, k_{2 \perp}^4} \approx -8 g^6 \, C_F \, \sigma_1 \, \sigma_2 \, \frac{s}{k_{1 \perp}^2 \, k_{2 \perp}^2}, \\
  & \langle |M|^2 \rangle \bigg|_C = 4 g^6 \, C_F \, \sigma_1 \, \sigma_2 \, s \, \frac{\ul{k}_2 \cdot (\ul{k}_1 - 2 \ul{k}_2)}{k_{1 \perp}^2 \, k_{2 \perp}^4} \approx -8 g^6 \, C_F \, \sigma_1 \, \sigma_2 \, \frac{s}{k_{1 \perp}^2 \, k_{2 \perp}^2}, \\
  & \langle |M|^2 \rangle \bigg|_D = 4 g^6 \, \frac{C_F}{N_c^2} \, \sigma_1 \, \sigma_2 \, s \, \frac{k_{1 \perp}^2 + k_{2 \perp}^2 + (\ul{k}_1 - \ul{k}_2)^2}{k_{1 \perp}^2 \, k_{2 \perp}^2 \, (\ul{k}_1 - \ul{k}_2)^2}, \\
  & \langle |M|^2 \rangle \bigg|_E = 4 g^6 \, \frac{C_F \, (N_c^2 -2)}{N_c^2} \, \sigma_1 \, \sigma_2 \, s \, \frac{k_{1 \perp}^2 + k_{2 \perp}^2 + (\ul{k}_1 - \ul{k}_2)^2}{k_{1 \perp}^2 \, k_{2 \perp}^2 \, (\ul{k}_1 - \ul{k}_2)^2}, \\
  & \langle |M|^2 \rangle \bigg|_F = 0, \\
  & \langle |M|^2 \rangle \bigg|_G = 0, \\
  & \langle |M|^2 \rangle \bigg|_H = 0, \\
  & \langle |M|^2 \rangle \bigg|_I = 4 g^6 \, \frac{C_F}{N_c^2} \,
  \sigma_1 \, \sigma_2 \, s \, \frac{k_{1 \perp}^2 + k_{2 \perp}^2 +
    (\ul{k}_1 - \ul{k}_2)^2}{k_{1 \perp}^2 \, k_{2 \perp}^2 \,
    (\ul{k}_1 - \ul{k}_2)^2}.
\end{align}
\end{subequations}
Note that in arriving at Eqs.~\eqref{ampl} we have added the top-down
and left-right mirror images of diagrams $B$, $C$, $E$, $H$, and the up-down
mirror images of $D$, $F$, $G$, $I$, as these transformations generate new
diagrams. The approximate expressions for diagrams $B$ and $C$ are
obtained by keeping their DLA contributions only: in such
contributions, the integration over the angles of $\ul{k}_1$ and/or
$\ul{k}_2$ eliminates the first term in the initial expressions for $B$
and $C$. Note that, after the extraction of these DLA contribution,
diagrams $B$ and $C$ cancel the diagram $A$,
\begin{align}
  \label{ABC}
  \langle |M|^2 \rangle \bigg|_A + \langle |M|^2 \rangle \bigg|_B +
  \langle |M|^2 \rangle \bigg|_C =0.
\end{align}
This is in complete analogy with the unpolarized (BFKL) case. Namely,
if we keep the leading-energy polarization-independent contributions
of the diagrams in \fig{Feyn_gauge23}, then \eq{ABC} would still hold.

The sum of all the diagrams in \fig{Feyn_gauge23} is then given by the
contributions of the bremsstrahlung diagrams $D$ -- $I$. We get
\begin{align}
  \label{sumAI}
  \langle |M|^2 \rangle \bigg|_A + \langle |M|^2 \rangle \bigg|_B +
  \ldots + \langle |M|^2 \rangle \bigg|_I = 4 g^6 \, C_F \, \sigma_1
  \, \sigma_2 \, s \, \frac{k_{1 \perp}^2 + k_{2 \perp}^2 + (\ul{k}_1
    - \ul{k}_2)^2}{k_{1 \perp}^2 \, k_{2 \perp}^2 \, (\ul{k}_1 -
    \ul{k}_2)^2}.
\end{align}

At this point it is appropriate to compare these results with the
discussion of Fig.~7 in \cite{Bartels:1996wc}. Our diagrams $B$ and $C$
from \fig{Feyn_gauge23} can be identified with the diagrams (d) and
(c) in Fig.~7 of \cite{Bartels:1996wc}, respectively, if one discards
the virtual photon lines and the upper quark propagator in the
latter. The discussion following Eq~(2.32) and continuing until the
end of Sec.~2 in \cite{Bartels:1996wc} also notes that \eq{ABC} holds
for the BFKL case, but appears to suggest that for helicity evolution
\eq{ABC} does not hold, and, instead, one has
\begin{align}
  \label{BC}
  \langle |M|^2 \rangle \bigg|_B + \langle |M|^2 \rangle \bigg|_C
  \stackrel{BER}{=} 0.
\end{align}
The conclusion \eqref{BC} was reached in \cite{Bartels:1996wc} for the
contribution of graphs $B$ and $C$ coming from the $k' = |\ul{k}_1 -
\ul{k}_2| \gg k_1$ regime, which seems to be identical to $k_2 \gg
k_1$ region of phase space, in which diagrams $B$ and $C$ are DLA. Guided
by \eq{BC}, the authors of \cite{Bartels:1996wc} conclude that
diagrams $B$ and $C$ cancel in the $k_2 \gg k_1$ regime, and need to be
considered in the $k' = |\ul{k}_1 - \ul{k}_2| \ll k_1, k_2$ region
only, where Gribov's theorem \cite{Gribov:1966hs} applies.

The conclusion \eqref{BC} reached in \cite{Bartels:1996wc} seems to
contradict the results of a direct calculation presented above in
Eqs.~\eqref{ampl} and resulting in \eq{ABC}. Note that our diagrams $B$
and $C$ come in with the same overall sign: this is due to their color
factors being different by a minus sign along with another minus sign
coming from the difference between the quark propagators to the right
of the cut. If \eq{BC} is incorrect, it appears the infra-red
evolution equations (IREE) derived in \cite{Bartels:1996wc} would need
to be modified, though we do not quite see how they could be changed
to easily accommodate the contributions of non-ladder graphs $B$ and $C$
in the $k_2 \gg k_1$ regime, where Gribov's theorem does not
apply. Note also that diagrams $D$, $E$ and $I$ give DLA contributions in
the same $k_2 \gg k_1$ regime, which do not cancel the contributions
of $B$ and $C$ from the same region of the phase space: it appears that
such contributions of $D$, $E$ and $I$ were not discussed in
\cite{Bartels:1996wc}, and they seem not to be taken into account by
the resulting IREE.

It is also possible that we misunderstood the discussion in
\cite{Bartels:1996wc} and the conclusion there was not given by
\eq{BC}, but, rather, the conclusion was that each diagram $B$ and $C$
separately is not DLA in the $k_2 \gg k_1$ regime. Unfortunately this
also seems to contradict the results of our calculations above in
Eqs.~\eqref{ampl}, which show that the contributions of diagrams $B$
and $C$ have the same momentum dependence as that of diagram $A$, and
hence $B$ and $C$ are DLA in the $k_2 \gg k_1$ regime in question,
since $A$ is also DLA in this region.


\end{document}